\documentclass[a4paper,11pt]{article}
\pdfoutput=1
\usepackage{graphicx}
\usepackage{epsf,amsmath,bbold,amsfonts,stmaryrd}

\usepackage[utf8]{inputenc}
\usepackage{mathrsfs}
\usepackage{appendix}
\usepackage{amssymb}
\usepackage{float}
\usepackage{color}
\usepackage{cite}
\usepackage[colorlinks]{hyperref}
\hypersetup{pageanchor=false}
\usepackage{indentfirst}
\usepackage{url}
\usepackage{float}
\usepackage{caption}
\usepackage[numbers,square,comma,sort&compress,merge]{natbib}

\newcommand{\lyxaddress}[1]{
\par {\raggedright #1
\vspace{1.4em}
\noindent\par}
}

\date{}

\def\nn{\nonumber}

\def\be{\begin{equation}}
\def\ee{\end{equation}}

\def\bea{\begin{eqnarray}}
\def\eea{\end{eqnarray}}

\def\ba{\begin{array}}
\def\ea{\end{array}}

\def\bc{\begin{center}}
\def\ec{\end{center}}

\def\bl{\begin{flushleft}}
\def\el{\end{flushleft}}

\def\br{\begin{flushright}}
\def\er{\end{flushright}}

\def\bi{\begin{itemize}}
\def\ei{\end{itemize}}

\def\bt{\begin{tabular}}
\def\et{\end{tabular}}

\numberwithin{equation}{section}

\begin{document}

\title{Greybody factors for Spherically Symmetric Einstein-Gauss-Bonnet-de
Sitter black hole}

\author{Cheng-Yong Zhang$^1$\thanks{zhangcy0710@pku.edu.cn},
Peng-Cheng Li$^2$\thanks{wlpch@pku.edu.cn} and Bin Chen$^{1,2,3}$\thanks{bchen01@pku.edu.cn}}
\maketitle

\lyxaddress{\begin{center}
1. Center for High Energy Physics, Peking University, 5 Yiheyuan
Road, Beijing 100871, China \\
2. Department of Physics and State Key Laboratory of Nuclear Physics
and Technology, Peking University, 5 Yiheyuan Road, Beijing 100871,
China\\
3. Collaborative Innovation Center of Quantum Matter, 5 Yiheyuan Road,
Beijing 100871, China
\par\end{center}}
\begin{abstract}
We study the greybody factors of the scalar fields in spherically symmetric
Einstein-Gauss-Bonnet-de Sitter black holes in higher dimensions. We derive the greybody factors analytically
for both minimally and non-minimally coupled scalar
fields. Moreover, we discuss the dependence of the greybody factor on  various
parameters including the angular momentum number, the non-minimally coupling
constant, the spacetime dimension, the cosmological constant and the Gauss-Bonnet
coefficient in detail. We find that the non-minimal coupling may suppress the {greybody factor} and the Gauss-Bonnet coupling could enhance it, {but they both suppress the energy emission rate of Hawking radiation.}

\end{abstract}
\newpage
\section{Introduction}

The black holes obey the laws  of thermodynamics
\cite{Bardeen1973Laws}. This inspires Hawking's pioneering works on the thermal radiation of the black hole, the so-called Hawking
radiation \cite{Hawking1975,Hawking1976}. Though in the vicinity of the black
hole event horizon, the Hawking radiation is a blackbody radiation, determined by the temperature of the black hole,
it becomes a greybody radiation at the asymptotic region since the radiation has to transverse an effective
potential barrier. This effective
potential barrier is highly sensitive to the structure of the
black hole background. As a result, the Hawking radiation encodes important information about the black hole, including
its mass, its charge and its angular momentum. In general, the Hawking radiation
of macroscopic black holes is too small to be detected. However, the Hawking
radiation of the
microscopic black holes could be detectable \cite{Argyres1998D}. Especially the existence of extra
spacelike dimensions \cite{ArkaniHamed1998a,ArkaniHamed1998b,Randall1999a,Randall1999b} indicates that
the tiny black hole may be created at the particle colliders
\cite{Giddings2001LHC,Dimopoulos2001LHC,Landsberg2006LHC,Dai2007LHC,Kanti2008LHC}
or in high energy cosmic-ray interactions \cite{Feng2001Cosmic,Emparan2001Cosmic,Ringwald2001Cosmic,Anchordoqui2001Cosmic}.
Thus the associated Hawking radiation maybe observed at the TeV scale.
As a result, significant number of works about the Hawking radiation in higher
dimensional spacetime have been done. For more extensive references
one may consult the reviews \cite{Landsberg2003,Kanti2004D,Harmark2007d,Kanti2014Hawking}.

In the asymptotic flat spacetimes, it has been found that the greybody
factors for the waves of arbitrary spin and angular quantum number $l$
in any dimensions vanish in the zero-frequency limit \cite{Page1976,Das1996,Higuchi2001},
even for the non-minimally coupled scalar \cite{Chen2010}. In the presence of  a positive cosmological constant, the picture is different.
The greybody factors of the Schwarzschild-de Sitter (SdS) black holes
were studied both analytically and numerically in  $d$ dimensions
in \cite{Kanti2005}, and it was found that the $l=0$
greybody factor was not vanishing even
in the zero-frequency limit for a minimally coupled massless scalar (see also \cite{Brady1996}). This implies that the cosmological constant has an important effect on the greybody factor as it leads to the fully delocalization of the zero-modes such that there is
a finite probability  for the zero-modes to transverse the region between the event horizon
and the cosmological horizon \cite{Kanti2005}. The mass
of the scalar or a non-minimally coupling constant breaks this relation,
hence the greybody factors for arbitrary non-minimally coupled scalar
partial modes in 4-dimensional spacetime tend to zero in the infrared
limit \cite{Crispino2013}.

In this paper, we consider the spherically symmetric dS black hole in the Einstein-Gauss-Bonnet
(EGB) gravity\footnote{In the following, we simply call such solution the EGB-dS black hole. }. The EGB gravity is a special case of the Lovelock gravity which
is the natural generalization of general relativity in higher dimensions
\cite{Lovelock1971}. As the most general metric theory of the gravity whose equations of motion are only
the second order differential equations, the Lovelock gravity
is ghost free and thus is especially attractive in the higher-derivative gravity
theories. Among the Lovelock gravity theories, the simplest one is the EGB gravity, which adds a fourth-derivative Gauss-Bonnet (GB) term to the Einstein-Hilbert action
\begin{equation}
S_{G}=\frac{1}{16\pi G}\int d^{d}x\sqrt{-g}\left[R+\alpha(R_{\mu\nu\rho\sigma}R^{\mu\nu\rho\sigma}-4R_{\mu\nu}R^{\mu\nu}+R^{2})\right].
\end{equation}
 Here $\alpha$ is the Gauss-Bonnet coupling constant of dimension
$(\mbox{length})^{2}$ and $R$ is the Ricci scalar. The Gauss-Bonnet coupling term appears in the low energy effective action of the heterotic string theory\cite{Boulware}, where the coupling constant $\alpha$ is positive definite and inversely proportional
to the string tension.
Hence in this work we consider the case that $\alpha\geq0$. $G$ is the $d$-dimensional
Newton's constant. The Gauss-Bonnet term is jut a topological
term in $d=4$ spacetime and becomes nontrivial in $d>4$ spacetimes.
It has been pointed out that if the Planck scale is of order TeV,
as suggested in some extra-dimension models, the coupling constant
$\alpha$ could be measured by LHC through the detection of the spectrum of the Hawking
radiation  of the black hole \cite{Barrau2003GBLHC}. Thus it
is worth studying the greybody factor of the Hawking radiation of the GB black
hole in higher dimensions, from both theoretical viewpoint and phenomenological purposes. For scalar and graviton emissions, the numerical studies of  the GB black hole in an asymptotic flat spacetime
were carried out in \cite{Grain2005,Konoplya2010}. As mentioned in the last paragraph,
the positive cosmological constant has significant
effect on the greybody factor. In this paper we would like to compute the greybody factor
of the Hawking radiation of the EGB-dS black hole analytically and discuss  the
effects of various parameters, especially the GB coupling constant, on the radiation.

The analytical study of the greybody factor in the SdS black hole has been
well-developed.  The analytical study in \cite{Kanti2005} was limited
to the case of the lowest partial mode ($l=0$) and the low energy part ($\omega\to0$) of the
spectrum. A general expression for the greybody factor for arbitrary
partial modes of a minimally or non-minimally coupled scalar in higher-dimensional
SdS black hole was derived in \cite{Kanti2014SdS}. The authors in \cite{Kanti2014SdS}
found an appropriate radial coordinate that allows them to integrate
the field equations analytically and avoid the approximations on the metric tensor used in \cite{Harmark2007d,Crispino2013}. The
comparison of the analytical result with the numerical result
was done in \cite{RPappas2016}. For more  studies, see \cite{Gonzalez2010,Gonzalez2011,Jorge2014,Dong2015,RSporea2015,RSakalli2016,RAhmed2016,RPanotopoulos2016,RMiao2016,RKanti2016,RPanotopoulos2016b,RPappasb,RPanotopoulosc}.
Adopting a similar radial coordinate, we are able to derive the analytical
results for the greybody factors for arbitrary partial modes of a scalar
field in the EGB-dS black hole spacetime as well.

In section \ref{section2}, we give the general background of the EGB-dS black hole
and the corresponding equation of motion for the scalar field. In section
\ref{section3}, we derive the analytical expression of the greybody factor using the matching
method and discuss its low energy limit.  We analyze the effects of various parameters on the greybody factor in section \ref{section4} and the energy emission of Hawking
radiation in section \ref{energyemission}. We end with the conclusion and discussion in section \ref{conclusion}.

\section{Background}\label{section2}

The  metric for a spherically symmetric Einstein-Gauss-Bonnet-de Sitter
black hole in $d$-dimensional spacetime is given by \cite{Boulware}
\begin{align}
ds^{2} & =-hdt^{2}+\frac{dr^{2}}{h}+r^{2}d\Omega_{d-2}^{2},\label{eq:4metric}\\
h & =1+\frac{r^{2}}{2\tilde{\alpha}}\left(1-\sqrt{1+\frac{4\tilde{\alpha}m}{r^{d-1}}+\frac{8\tilde{\alpha}\Lambda}{(d-1)(d-2)}}\right).\nonumber
\end{align}
The parameter $m$ is related to the mass of the black hole $M$ by
$m=\frac{16\pi GM}{(d-2)\Omega_{d-2}}$. In terms of the
horizon radius $r_{h}$, $m$ can be expressed as
\begin{equation}
m=r_{h}^{d-3}\left(1+\frac{\tilde{\alpha}}{r_{h}^{2}}-\frac{2\Lambda r_{h}^{2}}{(d-1)(d-2)}\right).
\end{equation}
Here $\tilde{\alpha}$ is related to the GB coupling
constant $\alpha$ by $\tilde{\alpha}=\alpha(d-3)(d-4)$. In the limit $\tilde{\alpha}\to0$,
the metric returns to that of the SdS black hole.
GB constant has significant effect on the stability of GB black holes. Through perturbation analysis, it was found that the EGB-dS black holes are unstable in certain parameter region. In our discussions, the parameters  are restricted to the stable region given in \cite{Konoplya2008,Konoplya2016,Konoplya2017} and will
be chosen such that the spacetime always has two horizons, the black
hole horizon $r_{h}$ and the cosmological horizon $r_{c}$.

We consider a general scalar field  coupled to the the gravity non-minimally
\begin{equation}
S_{\Phi}=-\frac{1}{2}\int d^{d}x\sqrt{-g}[\xi\Phi^{2}R+\partial_{\mu}\Phi\partial^{\mu}\Phi].
\end{equation}
Here $\xi$ is the non-minimally coupling constant with $\xi=0$ corresponding to the minimally coupled case.
The equation of motion of the scalar field has the form
\be
\nabla_{\mu}\nabla^{\mu}\Phi=\xi R\Phi.\ee
 In a spherically symmetric  background, we may make ansatz
  \be
  \Phi=e^{-i\omega t}\phi(r)Y_{(d-2)}^{l}(\Omega), \ee
where $Y_{(d-2)}^{l}(\Omega)$ are spherical harmonics on $S^{d-2}$.
Then the angular part and the radial part are decoupled such that  the radial
equation becomes
\begin{equation}
\frac{1}{r^{d-2}}\frac{d}{dr}\left(hr^{d-2}\frac{d\phi}{dr}\right)+\left[\frac{\omega^{2}}{h}-\frac{l(l+d-3)}{r^{2}}-\xi R\right]\phi=0.\label{eq:RadialEq}
\end{equation}
 Introducing $u(r)=r^{\frac{d-2}{2}}\phi(r)$, we get
\begin{equation}
\frac{d^{2}u}{dr_{\star}^{2}}+(\omega^{2}-V(r_{\star}))u=0,
\end{equation}
 where $r_{\star}$ is the tortoise coordinate defined by $dr_{\star}=dr/h(r)$.
The effective potential reads
\begin{equation}
V(r_{\star})=h\left[\frac{l(l+d-3)}{r^{2}}+\xi R+\frac{d-2}{2r}h'+\frac{(d-2)(d-4)}{4r^{2}}h\right].
\end{equation}
 It is obvious that the effective potential vanishes at the two horizons.
Its height increases with the angular momentum number $l$. Fixing
the black hole horizon $r_{h}=1$, we can study the dependence of the profile of the
effective potential on the angular momentum number $l$, the spacetime
dimension $d$, the scalar coupling constant $\xi$, the cosmological constant
$\Lambda$ and the GB coupling constant $\alpha$.

\section{Greybody factor }\label{section3}

The radial equation (\ref{eq:RadialEq}) can not be solved analytically
over the whole space region. However, to read the greybody factor, it is not necessary to
solve the equation exactly. Instead, one can solve the equation in two regions
separately, namely near the black hole horizon and the cosmological
horizon regions, and then paste the solutions in the intermediate region.  In
this procedure,  the effect of  the cosmological
constant should be put under control in order to make the result as accurate as possible \cite{Kanti2014SdS}.

\subsection{Near the event horizon}

In the near event horizon region $r\sim r_{h}$, similar to the case of SdS, we perform the following
transformation
\begin{equation}
r\rightarrow f(r)=\frac{h}{1-\tilde{\Lambda}r^{2}},\quad\tilde{\Lambda}=-\frac{1}{2\tilde{\alpha}}\left(1-\sqrt{1+\frac{8\tilde{\alpha}\Lambda}{(d-1)(d-2)}}\right) \label{eq:EventTrans}.
\end{equation}
The new variable $f$
ranges from 0 to $1$ as $r$ runs from $r_{h}$ to the region $r\gg r_{h}$.
Its derivative satisfies
\begin{equation}
\frac{df}{dr}=\frac{1-f}{r}\frac{A(r)}{1-\tilde{\Lambda}r^{2}},
\end{equation}
with
\be
A(r)=-2+\frac{d-1}{2}\left(1+\frac{1}{\sqrt{1+\frac{4\tilde{\alpha}m}{(1+2\tilde{\alpha}\tilde{\Lambda})^{2}}\frac{1}{r^{d-1}}}}\right)(1-\tilde{\Lambda}r^{2}),
\ee
in which the mass can be expressed as
\be
m=r_{h}^{d-3}(1-\tilde{\Lambda}r_{h}^{2})\left[1+\frac{\tilde{\alpha}(1+\tilde{\Lambda}r_{h}^{2})}{r_{h}^{2}}\right].\ee
When $\tilde{\alpha}\to0$, it returns to the case of the SdS black hole,
namely $A_{SdS}=-2+(d-1)(1-\tilde{\Lambda}r^{2})$.

Using the new variable, the radial equation near the even horizon becomes
\begin{equation}
f(1-f)\frac{d^{2}\phi}{df^{2}}+(1-B_{h}f)\frac{d\phi}{df}+\left[-\frac{(\omega r_{h})^{2}}{A_{h}^{2}}+\frac{(\omega r_{h})^{2}}{A_{h}^{2}f}-\frac{\lambda_{h}(1-\tilde{\Lambda}r_{h}^{2})}{A_{h}^{2}(1-f)}\right]\phi=0.\label{eq:RadialEqEevent}
\end{equation}
in which
\begin{align}
B_{h}=  2-\frac{1-\tilde{\Lambda}r_{h}^{2}}{A_{h}^{2}}\left[(d-3)A_{h}+rA'(r_{h})\right], \quad \lambda_{h}=  l(l+d-3)+\xi R^{(h)}r_{h}^{2},
\end{align}
 where $A_{h}=A(r_{h})$ and $R^{(h)}=-h''+(d-2)\frac{-2rh'+(d-3)(1-h)}{r^{2}}\Big|_{r_{h}}$
is the Ricci scalar on the event horizon. In the derivation of this equation we have used the
approximation
\be
\frac{(\omega r_{h})^{2}}{A_{h}^{2}f(1-f)}\sim\frac{(\omega r_{h})^{2}(1-f)}{A_{h}^{2}f}=-\frac{(\omega r_{h})^{2}}{A_{h}^{2}}+\frac{(\omega r_{h})^{2}}{A_{h}^{2}f},
\ee
near the event horizon $f\sim0$.
The reason is that the solution of the original radial equation has cusps due to the poles of Gamma function, the unphysical behavior can be avoided by using this  approximation\footnote{We thank Pappas and Kanti for their correspondences on this point.
}.

 This is in fact a Fuchsian equation with three singularities $f=0,1,\infty$.
To be clearer, make a redefinition $\phi=f^{\alpha_{1}}(1-f)^{\beta_{1}}W(f)$,
Eq.(\ref{eq:RadialEqEevent}) becomes
\begin{equation}
f(1-f)\frac{d^{2}W}{df^{2}}+\Big[1+2\alpha_{1}-\left(2\alpha_{1}+2\beta_{1}+B_{h}\right)f\Big]\frac{dW}{df}-\frac{\omega^{2}r_{h}^{2}+A_{h}^{2}(\alpha_{1}+\beta_{1})(B_{h}+\alpha_{1}+\beta_{1}-1)}{A_{h}^{2}}W=0.\label{Wf}
\end{equation}
in which the coefficients are given by
\begin{align}
\alpha_{1}=  \pm i\frac{\omega r_{h}}{A_{h}},\quad
\beta_{1}=  \frac{1}{2}\left(2-B_{h}\pm\sqrt{(2-B_{h})^{2}+\frac{4\lambda_{h}(1-\tilde{\Lambda}r_{h}^{2})}{A_{h}^{2}}}\right).
\end{align}
 The solution of the differential equation (\ref{Wf}) is the standard hypergeometric function $F(a_{1},b_{1},c_{1},f)$ with parameters $a_{1},b_{1},c_{1}$ being
 \begin{align}
a_{1}= & \alpha_{1}+\beta_{1}+\frac{1}{2}\left(B_{h}-1+\sqrt{(1-B_{h})^{2}-\frac{4\omega^{2}r_{h}^{2}}{A_{h}^{2}}}\right),\nonumber \\
b_{1}= & \alpha_{1}+\beta_{1}+\frac{1}{2}\left(B_{h}-1-\sqrt{(1-B_{h})^{2}-\frac{4\omega^{2}r_{h}^{2}}{A_{h}^{2}}}\right),\\
c_{1}= & 1+2\alpha_{1}.\nonumber
\end{align}
Considering the relation between $\phi(f)$ and $W(f)$, near the event horizon
the radial function $\phi(f)$ has the following form
\begin{align*}
\phi_{H} & =A_{1}f^{\alpha_{1}}(1-f)^{\beta_{1}}F(a_{1},b_{1},c_{1},f)+A_{2}f^{-\alpha_{1}}(1-f)^{\beta_{1}}F(1+a_{1}-c_{1},1+b_{1}-c_{1},2-c_{1},f).
\end{align*}
where $A_{1,2}$ are the constant coefficients. Near the event
horizon,
\be
\phi_{H}\simeq A_{1}f^{\alpha_{1}}+A_{2}f^{-\alpha_{1}}, \quad \mbox{and} \quad f\propto e^{A_{h}r_{\star}/r_{h}}.
\ee
Imposing the ingoing boundary condition near the event horizon and choosing $\alpha_{1}=-i\frac{\omega r_{h}}{A_{h}}$,
we should set $A_{2}=0$. Furthermore, the convergence of the hypergeometric
function requires the real part $Re(c_{1}-a_{1}-b_{1})>0$. Thus we
have to take the $``-"$ branch of $\beta_{1}$.
In the end, the solution near the event horizon is of the form
\begin{equation}
\phi_{H}=A_{1}f^{\alpha_{1}}(1-f)^{\beta_{1}}F(a_{1},b_{1},c_{1},f).\label{eq:EventSol}
\end{equation}

\subsection{Near the cosmological horizon}
The solution in the near cosmological horizon region can be solved
similarly. The function $h$ in the metric can be approximated by \cite{Harmark2007d,Crispino2013,Kanti2014SdS}
\begin{equation}
h(r)=1-\tilde{\Lambda}r^{2}-\left(\frac{r_{h}}{r}\right)^{d-3}(1-\tilde{\Lambda}r_{h}^{2})\sim \tilde{h}=1-\tilde{\Lambda}r^{2}.
\end{equation}
$\tilde{h}$ ranges from 0, at $r=r_{c}$, to 1 as $r\ll r_{c}$. In the above
approximation, the larger $r_{c}$  or the smaller $\tilde{\Lambda}$
leads to more accurate results. The approximation also becomes more
accurate for a larger spacetime dimension $d$.

Making the change of variable $r\rightarrow \tilde{h}(r)$, near the cosmological
horizon, we have
\begin{equation}\label{Xh}
\tilde{h}(1-\tilde{h})\frac{d^{2}\phi}{d\tilde{h}^{2}}+\left(1-\frac{d+1}{2}\tilde{h}\right)\frac{d\phi}{d\tilde{h}}+\left[\frac{(\omega r_{c})^{2}}{4\tilde{h}}-\frac{l(l+d-3)}{4(1-\tilde{h})}-\frac{\xi R^{(c)}r_{c}^{2}}{4}\right]\phi=0,
\end{equation}
 where $R^{(c)}=-\tilde{h}''+(d-2)\frac{-2r\tilde{h}'+(d-3)(1-\tilde{h})}{r^{2}}\Big|_{r_{c}}$
is the Ricci scalar at $r_{c}$. After a replacement $\phi(\tilde{h})=\tilde{h}^{\alpha_{2}}(1-\tilde{h})^{\beta_{2}}X(\tilde{h})$,
we get
\begin{equation}
(1-\tilde{h})\tilde{h}\frac{d^{2}X}{d\tilde{h}^{2}}+[1+2\alpha_{2}-(2\alpha_{2}+2\beta_{2}+\frac{d+1}{2})\tilde{h}]\frac{dX}{d\tilde{h}}-\frac{2(\alpha_{2}+\beta_{2})(\alpha_{2}+\beta_{2}+d-1)+\xi R^{(c)}r_{c}^{2}}{4}X=0,
\end{equation}
 in which
\begin{align}
\alpha_{2}= \pm i\frac{\omega r_{c}}{2},\quad
\beta_{2}=  -\frac{d+l-3}{2}\quad\mbox{or}\quad\frac{l}{2}.
\end{align}
 The solution of the differential equation (\ref{Xh}) could be written in terms of the  hypergeometric functions as well. Therefore, around the cosmological horizon, the radial equation can be solved
by
\begin{equation}
\phi_{C}=B_{1}\tilde{h}^{\alpha_{2}}(1-\tilde{h})^{\beta_{2}}F(a_{2},b_{2},c_{2},\tilde{h})+B_{2}\tilde{h}^{-\alpha_{2}}(1-\tilde{h})^{\beta_{2}}F(1+a_{2}-c_{2},1+b_{2}-c_{2},2-c_{2},\tilde{h}),
\end{equation}
 with the parameters
\begin{align}
a_{2}= & \alpha_{2}+\beta_{2}+\frac{d-1+\sqrt{(d-1)^{2}-4\xi R^{(c)}r_{c}^{2}}}{4},\\
b_{2}= & \alpha_{2}+\beta_{2}+\frac{d-1-\sqrt{(d-1)^{2}-4\xi R^{(c)}r_{c}^{2}}}{4},\nonumber \\
c_{2}= & 1+2\alpha_{2}.\nonumber
\end{align}
 Here $B_{1,2}$ are  constant coefficients. The convergence of the hypergeometric
function requires $Re(c_{2}-a_{2}-b_{2})>0$ such that we have to take $\beta_{2}=-\frac{d+l-3}{2}$.

Since the effective potential vanishes at $r_{c}$, the solution is
expected to be comprised of the plane waves. Indeed, we have
\be
\phi_{C}=B_{1}e^{-i\omega r_{\star}}+B_{2}e^{i\omega r_{\star}}\ee
where $r_{\star}=\frac{1}{2}r_{c}\ln\frac{r/r_{c}+1}{r/r_{c}-1}$
is the tortoise coordinate near $r_{c}$. The first and second parts
correspond to the ingoing and outgoing waves, respectively. The sign in
$\alpha_{2}$ just interchanges the ingoing and outgoing waves. We
take $\alpha_{2}=i\frac{\omega r_{c}}{2}$ here. In contrast to what
happens at the black hole horizon, both the ingoing and outgoing waves
are now allowed. It is in fact their amplitudes that define the greybody
factor for the emission of the scalar fields by the back hole. The greybody
factor is given by
\begin{equation}
|\gamma_{\omega l}|^{2}=1-\left|\frac{B_{2}}{B_{1}}\right|^{2}.
\end{equation}

\subsection{Matching the solutions in the intermediate region}

Now we have the asymptotic solutions in the near event horizon region and the near cosmological
horizon region. In order to complete the solution, we must ensure
that the two asymptotic solutions, $\phi_{H}$ and $\phi_{C}$ can
be smoothly pasted at the intermediate region.

\subsubsection{Black hole horizon}

First let us consider the near black hole horizon solution. Due to the fact that
in the intermediate region $r\gg r_{h}$, the variable
$f\rightarrow1$, we can use the following relation for the hypergeometric
function
\begin{align}
F(a,b,c;f)= & \frac{\Gamma(c)\Gamma(c-a-b)}{\Gamma(c-a)\Gamma(c-b)}F(a,b,a+b-c+1;1-f)\\
 & +(1-f)^{c-a-b}\frac{\Gamma(c)\Gamma(a+b-c)}{\Gamma(a)\Gamma(b)}F(c-a,c-b,c-a-b+1;1-f)\nonumber
\end{align}
to shift the argument from $f$ to $1-f$.
For simplicity we consider
the case $\Lambda r_{h}^{2}\ll1$. Then in the region where $r\gg r_{h}$,
we have $A_{h}\simeq d-3.$ This is reasonable only if $\Lambda r^{2}\simeq r^{2}/r_{c}^{2}\ll1$.
For $r\gg r_{h}$, from (\ref{eq:EventTrans}) we have
\begin{equation}
h\rightarrow1-\tilde{\Lambda}r^{2}+\mathcal{O}\left(\frac{r_{h}^{d-3}}{r^{d-3}}\right).
\end{equation}
Then the Ricci scalar $R^{(h)}\rightarrow\frac{2d\Lambda}{d-2}$.
Thus if $\xi$ is not too big, the term $\xi R^{(h)}r_{h}^{2}\rightarrow\xi\frac{2d\Lambda r_{h}^{2}}{d-2}\ll1$
and can be omitted. Therefore,  we have $B_{h}\simeq1,\beta_{1}\simeq-\frac{l}{d-3}$.

Now we have
\begin{eqnarray}
1-f&\simeq&\left(1+\frac{\tilde{\alpha}}{r_{h}^{2}}\right)\left(\frac{r_{h}}{r}\right)^{d-3}\nn\\
\beta_{1}+c_{1}-a_{1}-b_{1}&\simeq&\frac{l+d-3}{d-3}.
\end{eqnarray}
In the
intermediate region $r\gg r_{h}$, the solution (\ref{eq:EventSol}) can be expanded into the form
\begin{align}
\phi_{H} & \simeq\Sigma_{2}r^{l}+\Sigma_{1}r^{-l-d+3}\label{eq:EventMatch}
\end{align}
where
\begin{align}
\Sigma_{1}= & A_{1}\frac{\Gamma(c_{1})\Gamma(a_{1}+b_{1}-c_{1})}{\Gamma(a_{1})\Gamma(b_{1})}\left(1+\frac{\tilde{\alpha}}{r_{h}^{2}}\right)^{\frac{l+d-3}{d-3}}r_{h}^{l+d-3},\\
\Sigma_{2}= & A_{1}\frac{\Gamma(c_{1})\Gamma(c_{1}-a_{1}-b_{1})}{\Gamma(c_{1}-a_{1})\Gamma(c_{1}-b_{1})}\left(1+\frac{\tilde{\alpha}}{r_{h}^{2}}\right)^{\frac{-l}{d-3}}r_{h}^{-l}.\nonumber
\end{align}
 Note that the aforementioned approximations are applicable only for
the expressions involving the factor $(1-f)$ and not for the parameters
in the Gamma function to increase the validity of the analytical results
\cite{Kanti2014SdS}.

\subsubsection{Cosmological horizon}

Now let us turn to the solution near the cosmological horizon. Similar to the treatment above, we may shift
the argument of the hypergeometric function from $\tilde{h}$ to $1-\tilde{h}$ since
for the intermediate region $\tilde{h}\to1$. We still work with a small cosmological
constant. In the region where $r\ll r_{c}$, we have
\begin{equation}
1-\tilde{h}\simeq\left(\frac{r}{r_{c}}\right)^{2}
\end{equation}
 and $\beta_{2}\simeq-\frac{l+d-3}{2},~\beta_{2}+c_{2}-a_{2}-b_{2}\simeq l/2$.
Following the similar procedure, we get
\begin{align}
\phi_{C} & \simeq(\Sigma_{3}B_{1}+\Sigma_{4}B_{2})r^{-(l+d-3)}+(\Sigma_{5}B_{1}+\Sigma_{6}B_{2})r^{l}\label{eq:CosmMatch}
\end{align}
 where
\begin{align}
\Sigma_{3}=\frac{\Gamma(c_{2})\Gamma(c_{2}-a_{2}-b_{2})}{\Gamma(c_{2}-a_{2})\Gamma(c_{2}-b_{2})}r_{c}^{l+d-3}, & & \Sigma_{4}=\frac{\Gamma(2-c_{2})\Gamma(c_{2}-a_{2}-b_{2})}{\Gamma(1-a_{2})\Gamma(1-b_{2})}r_{c}^{l+d-3},\\
\Sigma_{5}=\frac{\Gamma(c_{2})\Gamma(a_{2}+b_{2}-c_{2})}{\Gamma(a_{2})\Gamma(b_{2})}r_{c}^{-l}, & & \Sigma_{6}=\frac{\Gamma(2-c_{2})\Gamma(a_{2}+b_{2}-c_{2})}{\Gamma(a_{2}-c_{2}+1)\Gamma(b_{2}-c_{2}+1)}r_{c}^{-l}.\nonumber
\end{align}
It is obvious that solutions (\ref{eq:EventMatch}) and (\ref{eq:CosmMatch})
have the same power-law. Identifying the coefficients of the same
powers of $r$ in (\ref{eq:EventMatch}) and (\ref{eq:CosmMatch}),
we get the relations
\begin{equation}
\Sigma_{3}B_{1}+\Sigma_{4}B_{2}=\Sigma_{1},\quad \Sigma_{5}B_{1}+\Sigma_{6}B_{2}=\Sigma_{2}.
\end{equation}
Solving the constraints and plugging them into the expression for
the greybody factor for the emission of scalar fields by a higher
dimensional EGB-dS black hole, we get
\begin{equation}
|\gamma_{\omega l}|^{2}=1-\left|\frac{\Sigma_{2}\Sigma_{3}-\Sigma_{1}\Sigma_{5}}{\Sigma_{1}\Sigma_{6}-\Sigma_{2}\Sigma_{4}}\right|^{2}.\label{eq:GammaGrey}
\end{equation}
 This expression takes the same form as that for the Einstein gravity
\cite{Kanti2014SdS}. But due to the differences among the explicit expressions of $\Sigma$s, it depends not only on the cosmological constant $\Lambda$ and the non-minimal coupling $\xi$, but also on the GB coupling
constant $\alpha$.

As mentioned in \cite{Kanti2014SdS}, the greybody factor (\ref{eq:GammaGrey}) is more accurate
for a smaller cosmological constant and a larger distance between $r_{h}$
and $r_{c}$. On the other hand, in contrast with all the previous
similar matching procedures  here we do not make any assumption on the energy $\omega$
in the approximation, thus it might be possible that our analytical result can be valid beyond the low energy region. However, as
we will see  in the following section there are obvious deviations in the high energy region from the reasonable expected results, which means that the matching procedure only applies to the low energy region. This is because that  the continuations of the asymptotic solutions near the event/cosmological horizon deviate from the exact solution in the intermediate region so the higher energy modes lead to larger deviations. Instead, one can numerically integrate the radial equation (\ref{eq:RadialEq}) in the intermediate region to get the more exact greybody factors for high energy modes. We leave this to future work.

\subsection{Low energy limit}

As we mentioned above, the analytical result of the greybody factor is only valid for low energy modes, therefor before analyzing the effects of various parameters on the greybody
factor, we derive the low energy limit of the greybody factor in this
subsection.
\subsubsection{Minimal coupling $\xi=0$ and dominant mode $l=0$ }

Let us consider the minimally coupling $\xi=0$ case and the dominant
mode $l=0$ first. In this case,
we obtain
\begin{align}
\Sigma_{1}\sim A_{1}\frac{i\omega}{2-B_{h0}}\left(1+\frac{\tilde{\alpha}}{r_{h}^{2}}\right)\frac{1}{A_{h0}}r_{h}^{d-2}+O(\omega^{2}), & \quad \Sigma_{2}\sim A_{1}+O(\omega),\\
\Sigma_{3}\sim\frac{i\omega}{d-3}r_{c}^{d-2}+O(\omega^{2}),\quad \Sigma_{4}\sim\frac{-i\omega}{d-3}r_{c}^{d-2}+O(\omega^{2}), & \quad \Sigma_{5,6}\sim1+O(\omega)\nonumber
\end{align}
where

\begin{align}
A_{h0}= & \frac{(d-3)r_{h}^{2}+(d-5)\tilde{\alpha}}{r_{h}^{2}+2\tilde{\alpha}},\quad B_{h0}=\frac{(d-3)r_{h}^{2}-4\tilde{\alpha}}{(d-3)r_{h}^{2}+(d-5)\tilde{\alpha}},\\
\lambda_{h0}= & l(l+d-3)+(d-1)\tilde{\alpha}\xi\frac{(2-d)r_{h}^{4}+4r_{h}^{2}\tilde{\alpha}+2(d+1)\tilde{\alpha}^{2}}{(r_{h}^{2}+2\tilde{\alpha})^{3}}.\nonumber
\end{align}
 Then the greybody factor becomes
\begin{equation}
|\gamma_{\omega l}|^{2}=\frac{4(d-3)A_{h0}(2-B_{h0})\left(1+\frac{\tilde{\alpha}}{r_{h}^{2}}\right)(r_{h}r_{c})^{d-2}}{\left[(d-3)\left(1+\frac{\tilde{\alpha}}{r_{h}^{2}}\right)r_{h}^{d-2}+A_{h0}(2-B_{h0})r_{c}^{d-2}\right]^{2}}+O(\omega).\label{eq:GreyLow}
\end{equation}
 Thus the scalar particle with very low energy has a non-vanishing probability
of being emitted by a higher dimensional EGB-dS black hole. This is in fact
a characteristic feature of the propagation of free massless scalar
in the dS spacetime. However, the GB term changes the value of the greybody factor. For instance, for a small
$\tilde{\alpha}$, up to the first order of $\tilde{\alpha}$,
\begin{equation}
|\gamma_{\omega l}|^{2}=\frac{4(r_{h}r_{c})^{d-2}}{\left(r_{h}^{d-2}+r_{c}^{d-2}\right)^{2}}+\frac{4r_{c}^{d-2}r{}_{h}^{d-2}(r_{c}^{d-2}-r_{h}^{d-2})}{(r_{c}^{d-2}+r_{h}^{d-2})^{3}}\frac{\tilde{\alpha}}{r_{h}^{2}}+O(\omega,\tilde{\alpha}^{2}).
\end{equation}
We see that $\tilde{\alpha}$ increases the greybody factor of
massless scalar in the EGB-dS black hole background. When $\tilde{\alpha}\to0$,
we reproduce the low energy  greybody factor for the mode $l=0$, in accordance to the previous higher dimensional
analysis \cite{Harmark2007d,Kanti2005,Kanti2014SdS}.
\subsubsection{Non-minimally coupling case $\xi\protect\neq0$ }

Now we calculate the low energy greybody factor for a non-minimally coupled
scalar. In this case, we can expand the the combinations in the low energy limit as
\bea
\Sigma_{2}\Sigma_{3}=E+i\Sigma_{231}\omega+\Sigma_{232}\omega^{2}, & & \Sigma_{1}\Sigma_{5}=K+i\Sigma_{151}\omega+\Sigma_{152}\omega^{2},\\
\Sigma_{2}\Sigma_{4}=E+i\Sigma_{241}\omega+\Sigma_{242}\omega^{2}, & & \Sigma_{1}\Sigma_{6}=K+i\Sigma_{161}\omega+\Sigma_{162}\omega^{2}.\nonumber
\eea
 in which $E,K,\Sigma_{231},\Sigma_{232},\Sigma_{151},\Sigma_{152},\Sigma_{241},\Sigma_{242},\Sigma_{161},\Sigma_{162}$
are the expansion coefficients, whose explicit expressions are lengthy and will not be given here. The final result for the greybody
factor turns out to be
\begin{equation}
|\gamma_{\omega l}|^{2}= \frac{4\pi^{8}(r_{c}r_{h})^{d+2l+3}R_{C}R_{H}\left(1+\frac{\tilde{\alpha}}{r_{h}^{2}}\right)^{\frac{2l+d+3}{d-3}}}{\sin^{2}(\pi\delta)\sin^{2}(\pi\epsilon)\sin^{2}(\pi\eta_{+})\sin^{2}(\pi\eta_{-})(\delta+\epsilon-1)(\eta_{+}+\eta_{-}-1)C^{2}}\omega^{2}+O(\omega^{3}),\\
\end{equation}
 in which $R_{H}=\frac{r_{h}}{A_{h0}},R_{C}=\frac{r_{c}}{2}$ and
\bea
\delta&= & \frac{1}{2}\left(B_{h0}-\sqrt{(2-B_{h0})^{2}+\frac{4\lambda_{h0}}{A_{h0}^{2}}}\right),\nn\\
\epsilon&=&\frac{1}{2}\left(2-B_{h0}-\sqrt{(2-B_{h0})^{2}+\frac{4\lambda_{h0}}{A_{h0}^{2}}}\right),\nonumber \\
\eta_{\pm}&= & \frac{5-d-2l\pm\sqrt{(d-1)^{2}-4\xi R^{(c)}r_{c}^{2}}}{4},\label{eq:LowPara}
\eea
 and
\begin{align}
C= & r_{c}^{3}r_{h}^{d+2l}\left(1+\frac{\tilde{\alpha}}{r_{h}^{2}}\right)^{\frac{d+2l}{d-3}}\Gamma(1-\delta)\Gamma(1-\epsilon)\Gamma(\delta+\epsilon-1)\Gamma(1-\eta_{+})\Gamma(1-\eta_{-})\Gamma(\eta_{+}+\eta_{-}-1)\nonumber \\
 & -r_{c}^{d+2l}r_{h}^{3}\left(1+\frac{\tilde{\alpha}}{r_{h}^{2}}\right)^{\frac{3}{d-3}}\Gamma(\delta)\Gamma(\epsilon)\Gamma(1-\delta-\epsilon)\Gamma(\eta_{+})\Gamma(\eta_{-})\Gamma(1-\eta_{+}-\eta_{-}).\nonumber
\end{align}

 Note that the first non-vanishing term in the low energy expansion
is of order $O(\omega^{2})$. This holds for all partial waves including
the dominant mode $l=0$. Therefore, there is no mode with a non-vanishing
low energy greybody factor for the non-minimally coupled scalar. This has a simple explanation: from
the equation of motion for the non-minimally coupled scalar, we see
that the coupling constant $\xi$ plays a role of an effective mass
 for the scalar and breaks the infrared enhancement, as mentioned
in the introduction.

{\footnotesize{}}
\begin{figure}[h]
\begin{centering}
{\footnotesize{}}%
\begin{tabular}{cc}
{\footnotesize{}\includegraphics[scale=0.6]{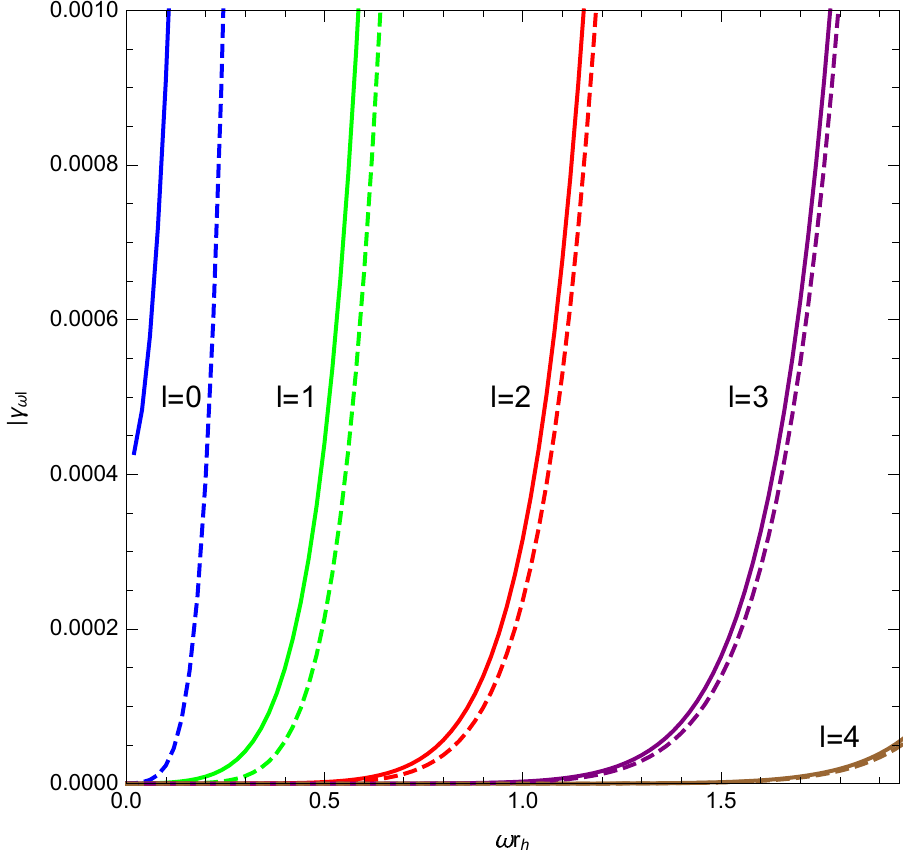}} & {\footnotesize{}\includegraphics[scale=0.6]{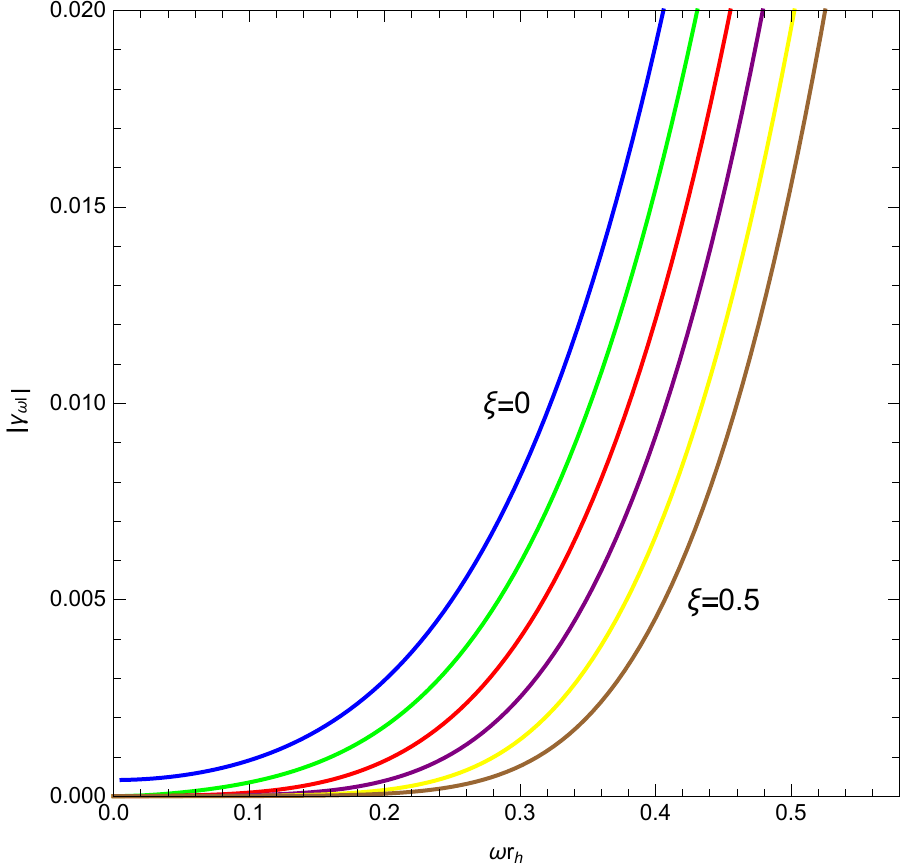}}\tabularnewline
{\footnotesize{}\includegraphics[scale=0.6]{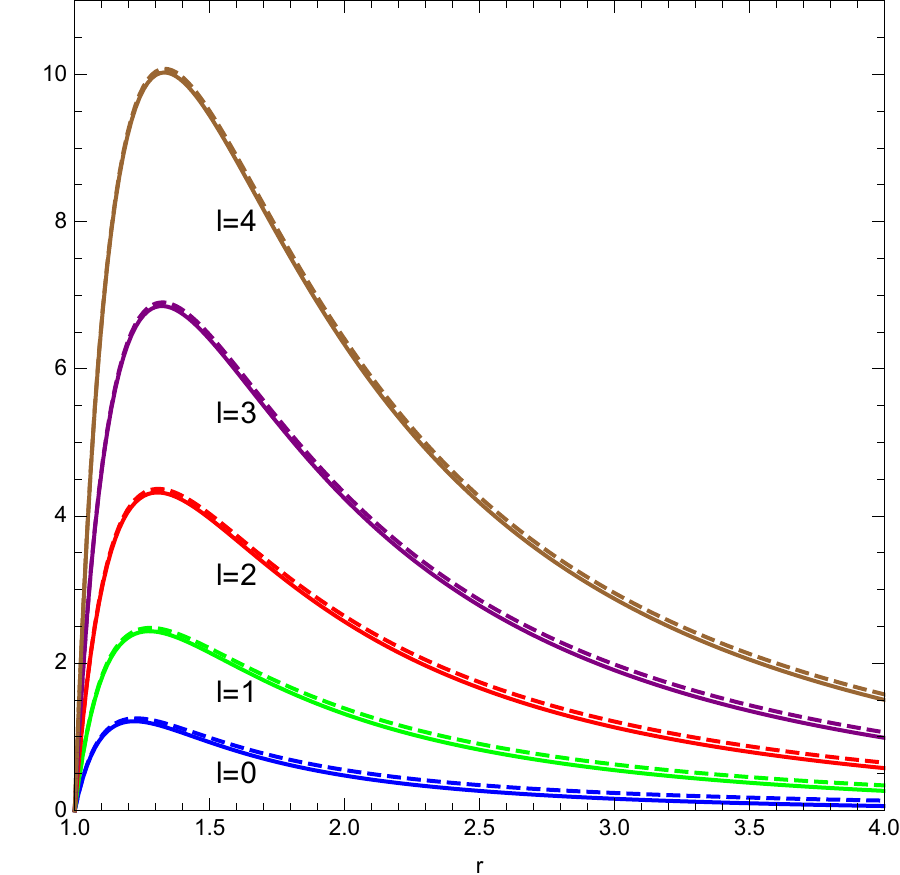}} & {\footnotesize{}\includegraphics[scale=0.6]{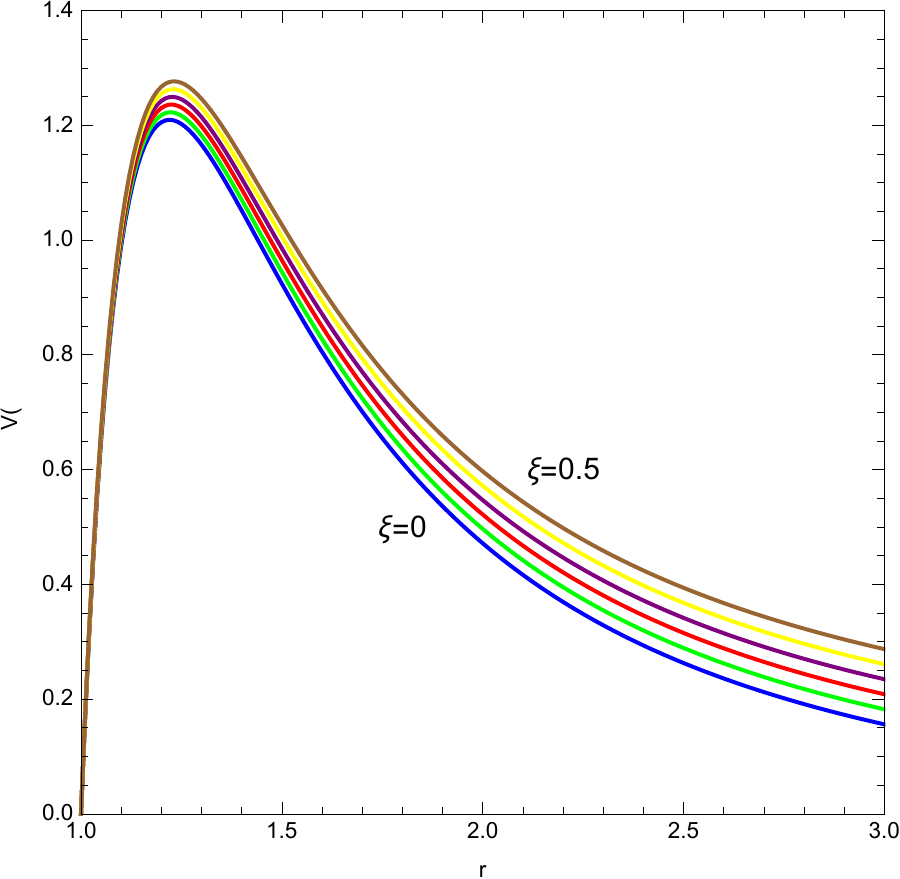}}\tabularnewline
\end{tabular}
\par\end{centering}{\footnotesize \par}
{\footnotesize{}\caption{\label{fig:GreyLxi} Effects of parameters $l$ and $\xi$. The greybody factors (upper panel) and corresponding
effective potential (lower panel) for the scalar fields when $d=6,\Lambda=0.1,\tilde{\alpha}=0$.
Left panel for $l=0,1,2,3,4$ and $\xi=0$ (solid lines) or $\xi=0.3$
(dashed lines). Right panel for $l=0$ and $\xi=0,0.1,0.2,0.3,0.4,0.5$.}
}{\footnotesize \par}
\end{figure}

\section{The effects of various parameters}\label{section4}

There are several parameters in the theory which influence
the greybody factor for the non-minimally coupled scalar propagating in
the EGB-dS black hole spacetime. These parameters include the non-minimally coupling constant $\xi$, angular momentum number $l$, the spacetime dimension $d$, the cosmological constant $\Lambda$ and the GB coupling constant $\tilde{\alpha}$.
 In fact, the parameters $\xi,l,d,\Lambda$ have the similar effects on the greybody factor of the EGB-dS black hole as they have for that of the SdS black hole.  Therefore, we focus on the effect of the GB coupling constant $\tilde{\alpha}$ on the greybody factor.
To analyze their effects more clearly, we plot the dependence of the greybody
factor on these parameters and the corresponding effective potentials in the following.

\subsection{The case $\tilde{\alpha}=0$}\label{section4.1}
For the purpose of comparison, we produce Fig.\ref{fig:GreyLxi} to show that our results agree with the SdS results (figure 8 in \cite{Kanti2014SdS}) in the limit $\tilde{\alpha}\to 0$ . From Fig.\ref{fig:GreyLxi} we see that the
suppression of the greybody factor by the angular momentum number $l$ is
obvious in the left upper panel, both for minimally or non-minimally coupled
scalar. As shown in (\ref{eq:GreyLow}), for the dominant mode $l=0$
of the minimally coupled scalar $\xi=0$, we find a non-vanishing greybody
factor for the low energy emission. While for the non-minimally coupled
scalar, the greybody factor for the low energy mode vanishes. Moreover,
$\xi$ decreases the greybody factor when other parameters
are fixed. We plot the effective potential in the lower panel to have
an intuitive explanation. It can be seen that the effective potential barriers become higher with $\xi$, as a consequence it becomes
more difficult for the scalar to transverse the barrier to reach the
near horizon region. So the greybody factor  decreases with $\xi$.

\begin{figure}[h]
\begin{centering}
\begin{tabular}{cc}
\includegraphics[scale=0.6]{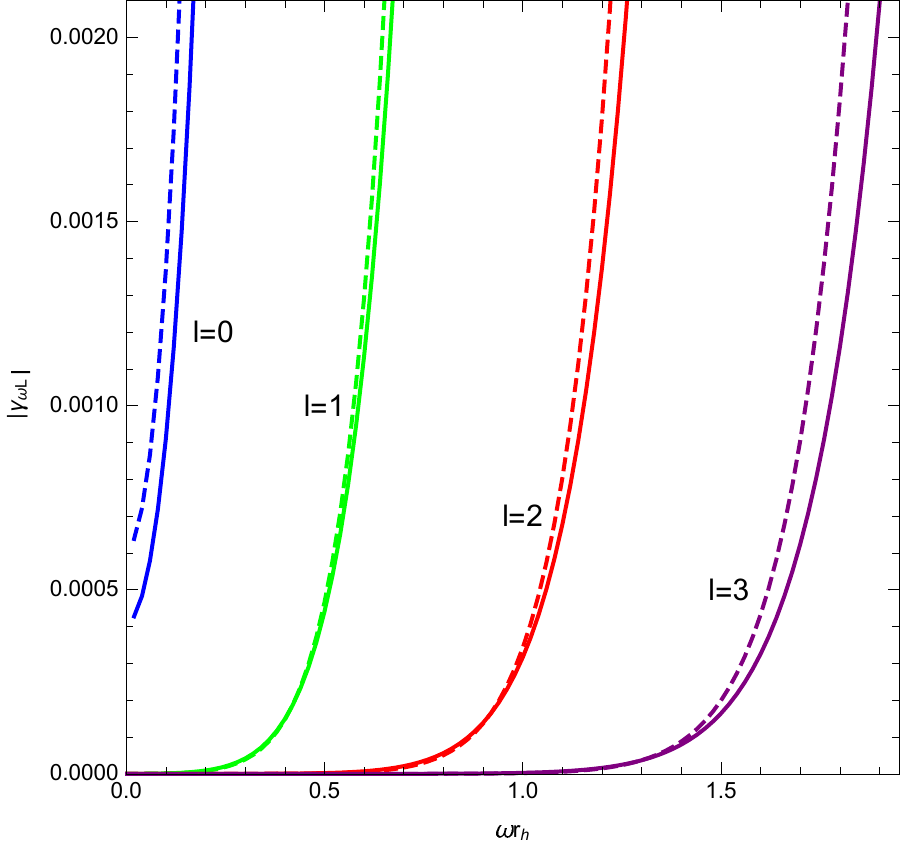} & \includegraphics[scale=0.6]{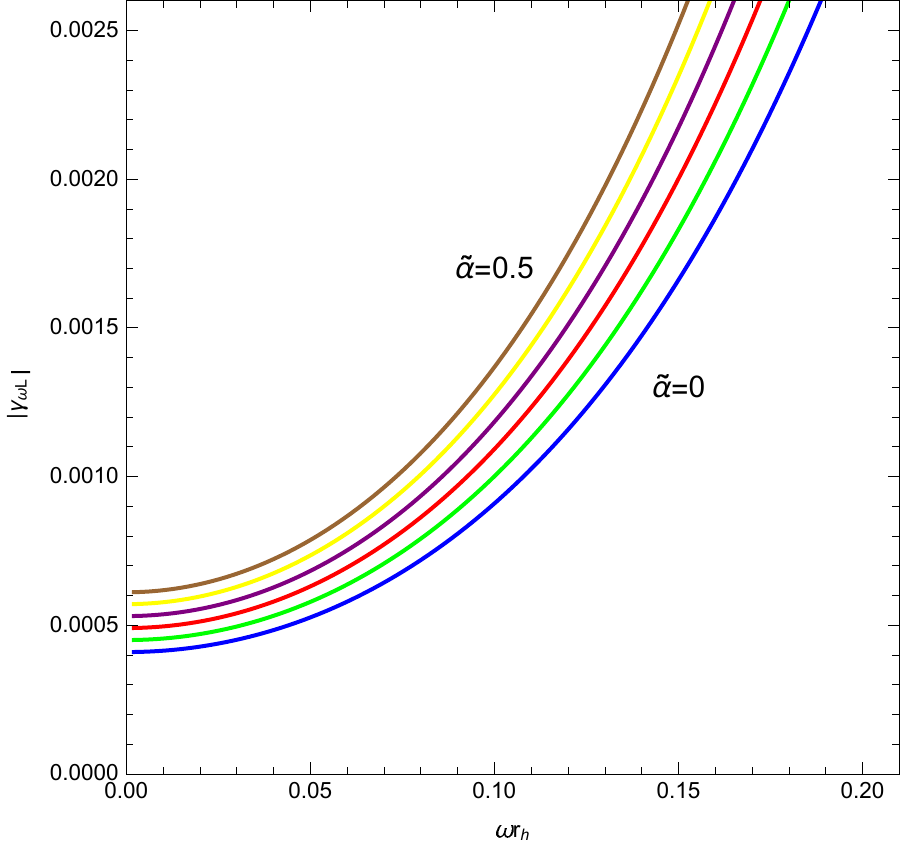}\tabularnewline
\includegraphics[scale=0.6]{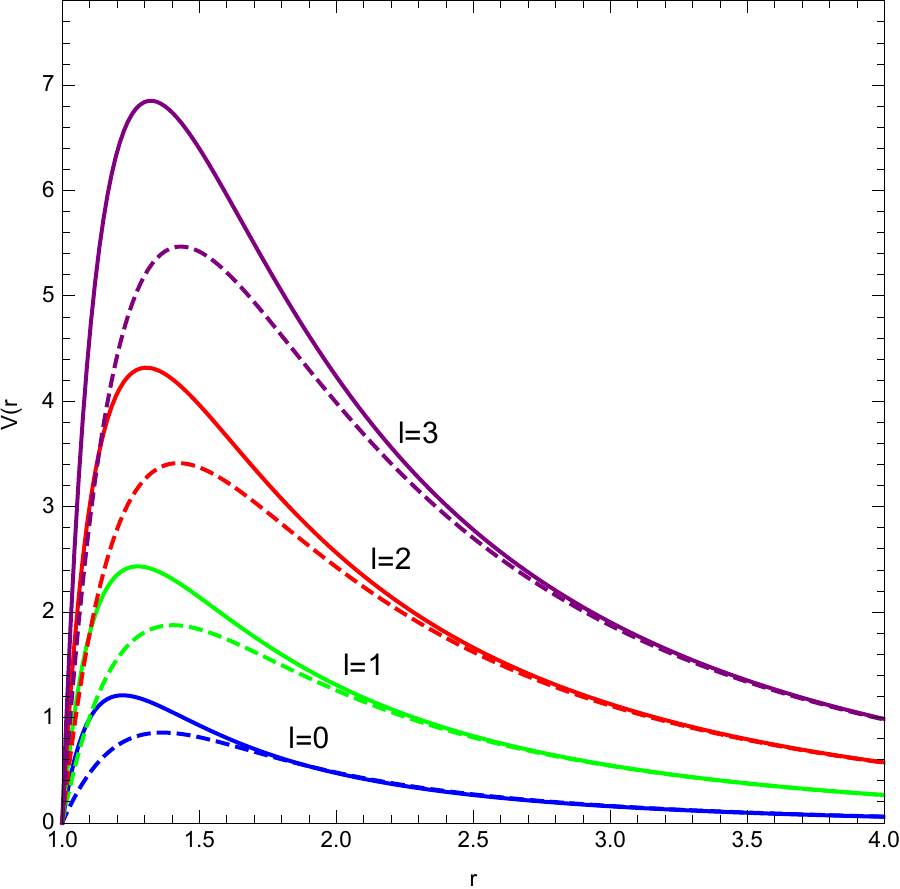} & \includegraphics[scale=0.6]{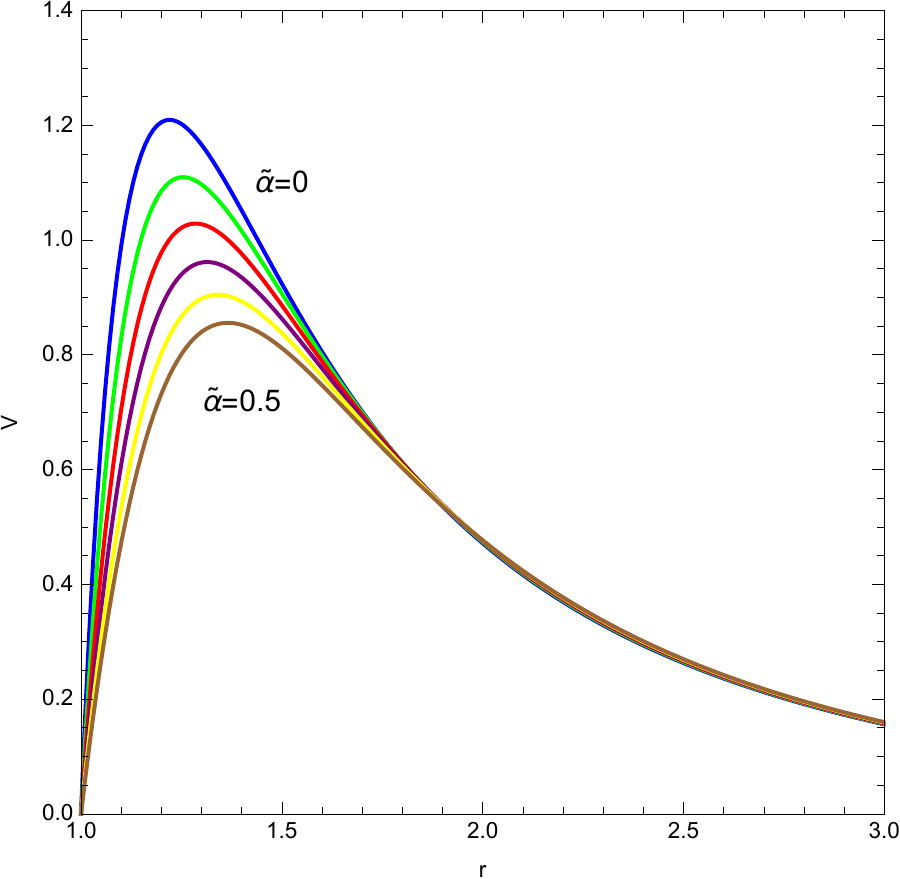}\tabularnewline
\end{tabular}
\par\end{centering}
\caption{\label{fig:GreyAlpha} Effects of $l$ and $\tilde{\alpha}$. The greybody factors (upper panel) and corresponding
effective potential (lower panel) for $d=6,\Lambda=0.1,\xi=0$. Left
panel for $l=0,1,2,3$ and $\tilde{\alpha}=0$ (solid lines) or $\tilde{\alpha}=0.5$
(dashed lines). Right panel for $l=0$ and $\tilde{\alpha}=0,0.1,0.2,0.3,0.4,0.5$.}
\end{figure}

\subsection{Effects of $\tilde{\alpha}$}

Now we study the effects of the Gauss-Bonnet parameter $\tilde{\alpha}$ on the greybody factor.

\subsubsection{Effects of $\tilde{\alpha}$ on different partial modes $l$}

In Fig.\ref{fig:GreyAlpha} we plot the greybody factor for the minimally coupled scalar when  $\tilde{\alpha}=0.5$. From the
left upper panel, we find the suppression of the greybody factor by
the angular momentum number $l$ as well. For the dominant mode $l=0$, there
is a non-vanishing greybody factor for the low energy modes. Unlike the
case that the greybody factors for zero modes  vanish when $\xi\neq0$,
the presence of $\tilde{\alpha}$ makes it have a non-zero value.
The greybody factors with respect to $\tilde{\alpha}$ for the
dominant mode are shown in the right upper panel. It is obvious that the greybody
factor does not vanish when $\omega=0$. Actually, it increases with
$\tilde{\alpha}$ . We plot the corresponding effective potential
in the lower panel to give an intuitive interpretation. The effective
potential decreases with $\tilde{\alpha}$ when other parameters are
fixed. Thus it becomes easier for the scalar to transverse it and
the greybofy factor is enhanced with $\tilde{\alpha}$.

\subsubsection{The competition between  $\tilde{\alpha}$  and $\xi$ }

\begin{figure}[h]
\begin{centering}
\begin{tabular}{cc}
\includegraphics[scale=0.7]{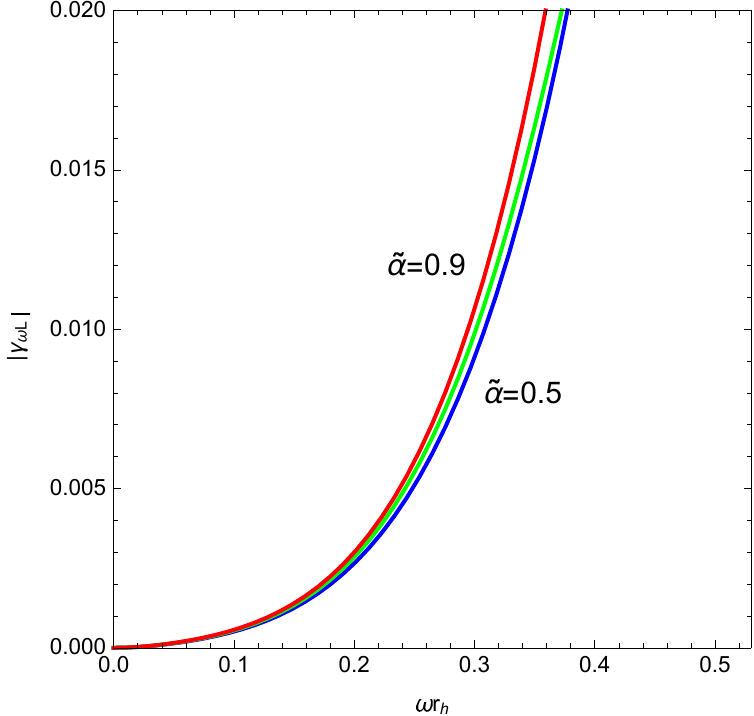} & \includegraphics[scale=0.7]{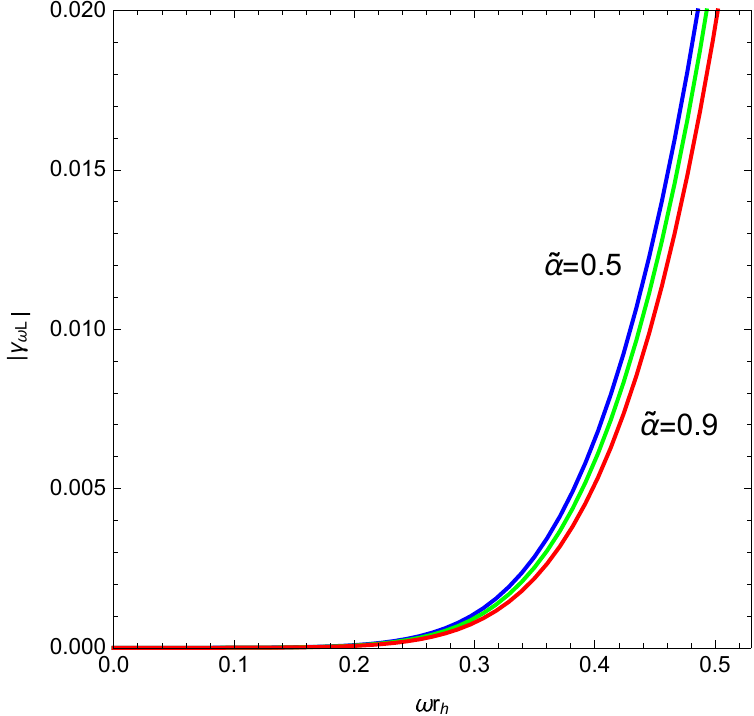}\tabularnewline
\end{tabular}
\par\end{centering}
\caption{\label{fig:GreyXiAlpha} The competition between $\xi$ and $\tilde{\alpha}$. Greybody factors for $d=6,l=0,\Lambda=0.1$,
$\xi=0.1$ (left) and $\xi=0.5$ (right) with respect to $\tilde{\alpha}=0.5,0.7,0.9$
respectively.}
\end{figure}

Since the non-minimally coupling $\xi$ suppresses the Hawking radiation (as we can see in section \ref{section4.1}) while the Gauss-Bonnet term $\tilde{\alpha}$
enhances it, there must be a competition between them. In Fig.\ref{fig:GreyXiAlpha},
we find that when $\xi$ is small ($\xi=0.1$ in the left panel),
 $\tilde{\alpha}$ increases the greybody factor. When
$\xi$ is large ($\xi=0.5$ in the right panel), $\tilde{\alpha}$
decreases the greybody factor. This phenomenon appears also for $\xi$
and $\Lambda$ which will be shown in subsection \ref{section4.4}. However, unlike
the competition between $\xi$ and $\Lambda$, the competition between
$\xi$ and $\tilde{\alpha}$ is too involved for us to have an intuitive
analysis from the effective potential.

\subsubsection{Effects of  $\tilde{\alpha}$ on modes in different dimensional spacetimes $d$}

\begin{figure}[h]
\begin{centering}
\begin{tabular}{cc}
\includegraphics[scale=0.3]{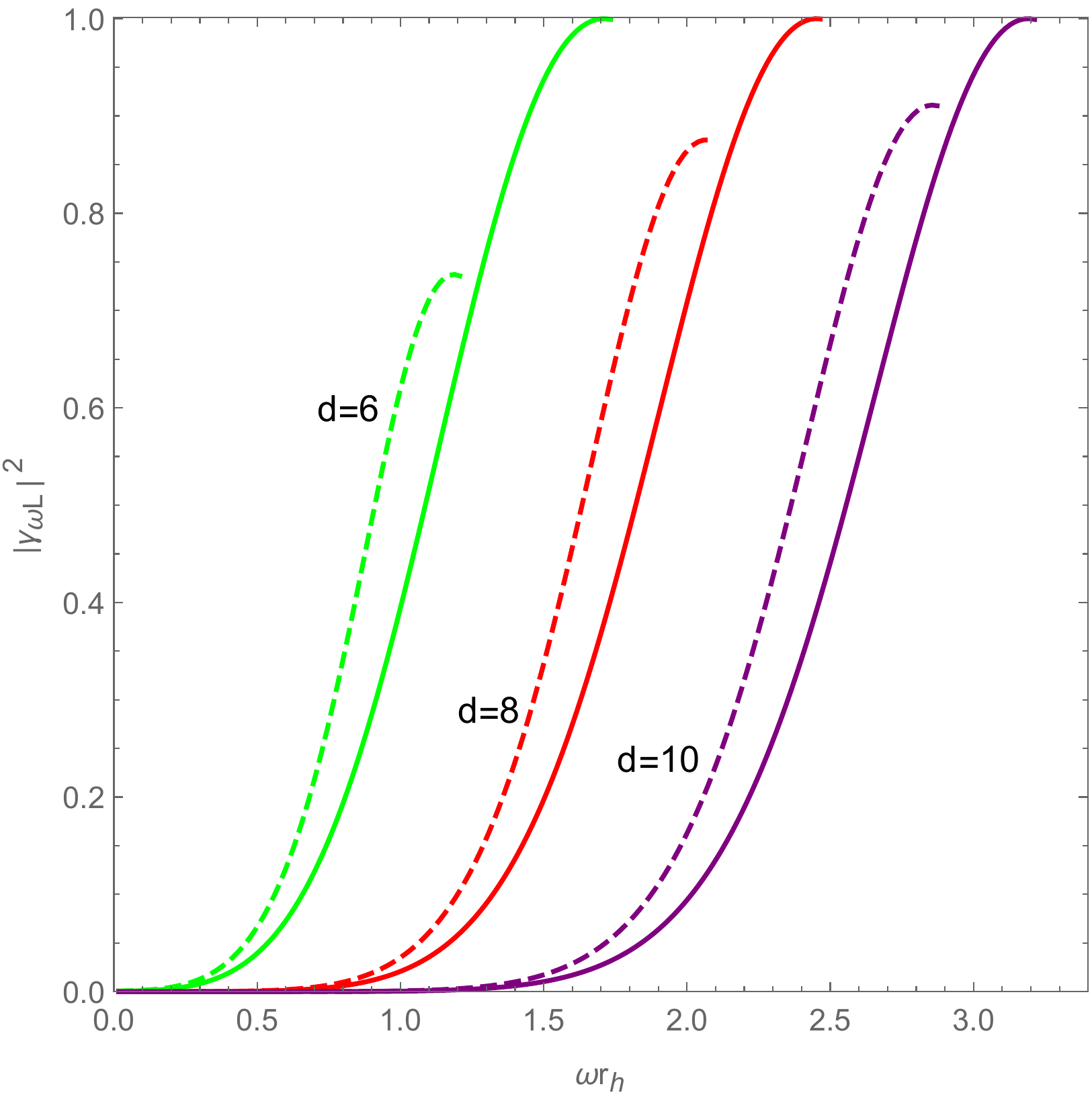} & \includegraphics[scale=0.3]{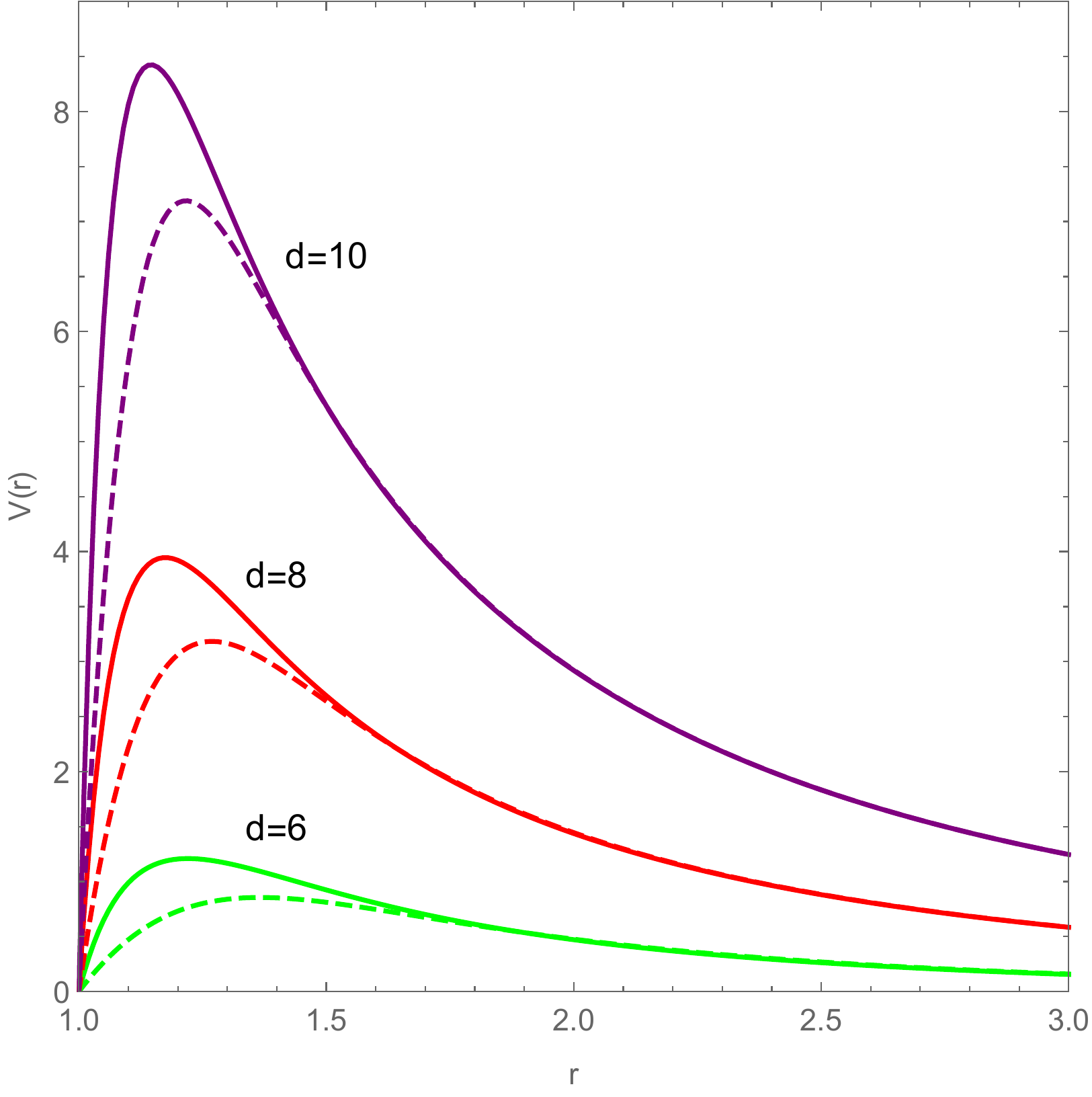}\tabularnewline
\end{tabular}
\par\end{centering}
\caption{\label{fig:GreyD} Effects of $d$ and $\tilde{\alpha}$. The greybody factors (left panel) and corresponding
effective potentials (right panel) for $\Lambda=0.1,\xi=0,l=0$ and
$d=6,8,10$ with $\tilde{\alpha}=0$ (solid lines) or $\tilde{\alpha}=0.5$
(dashed lines). }
\end{figure}

Now let us study the dependence of the greybody factor on the spacetime
dimension $d$ in the presence of  $\tilde{\alpha}$. In Fig.\ref{fig:GreyD}, we see that the greybody
factor is significantly suppressed in higher dimensions. For example, for
$d=6,8,10$ the greybody factors for the minimally coupled scalar at $\omega=0$ have values of order $10^{-4},10^{-7}$
and $10^{-10}$, respectively. For different dimensions the greybody factor still increases with $\tilde{\alpha}$.
We plot the
effective potential in the right panel. We see that the potential
barrier increases significantly with $d$. Thus it becomes harder
for the scalar to transverse the barrier and the greybody factor decreases
with $d$. On the other hand,  $\tilde{\alpha}$ decreases
the potential barrier and so increases the
greybody factor \footnote{For the large $d$ behavior of the EGB black holes, one can find the study in \cite{Chen1511,Chen1703}.}.

Note that in Fig.\ref{fig:GreyD} we plot only the greybody factors in the low energy region. In the high energy region the greybody factors decrease to zero which is unreasonable since the high energy modes can transverse the potential barrier easier and the
greybody factors  should approach to 1. Thus as we mentioned before, though we do not restrict energy $\omega$ in the derivation of the greybody factors, this matching approach is still limited to low energy region.
Moreover, due to the poles of the Gamma functions in the solution,
we are not able to obtain the analytical results for odd dimensional spacetimes. A complete analysis is needed and we leave it to future work.

\subsubsection{\label{subsec:Competition}Competition between $\xi$ and $\Lambda$ in the presence of $\tilde{\alpha}$}\label{section4.4}

\begin{figure}[h]
\begin{centering}
\begin{tabular}{cc}
\includegraphics[scale=0.6]{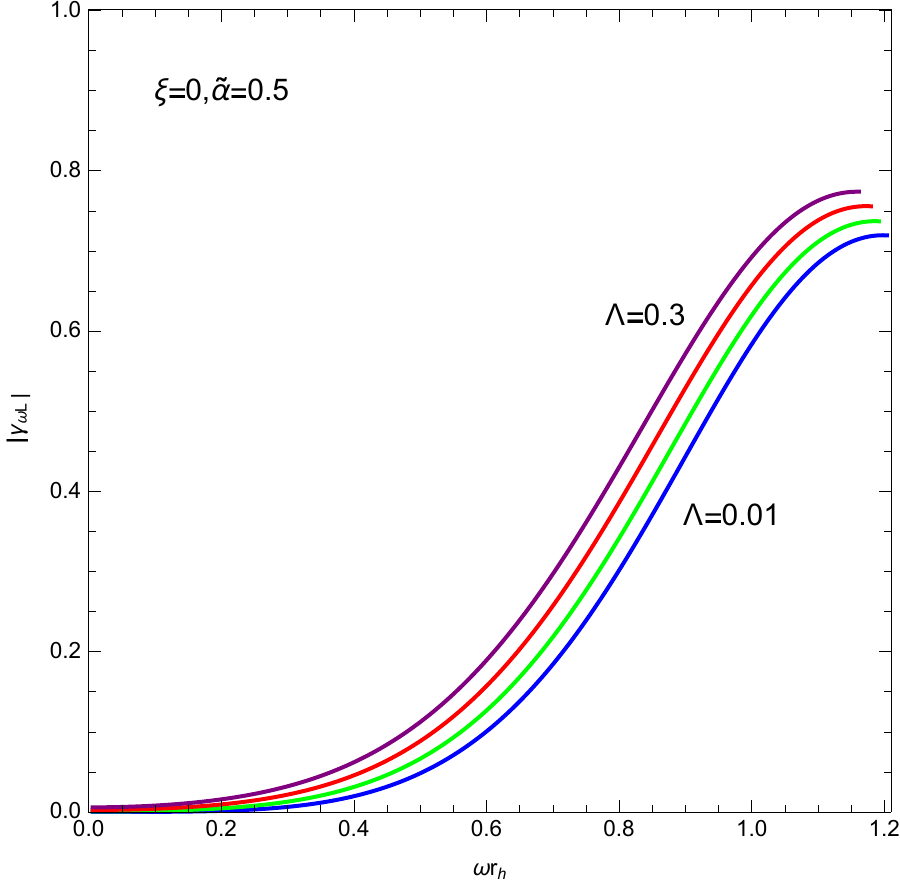} & \includegraphics[scale=0.6]{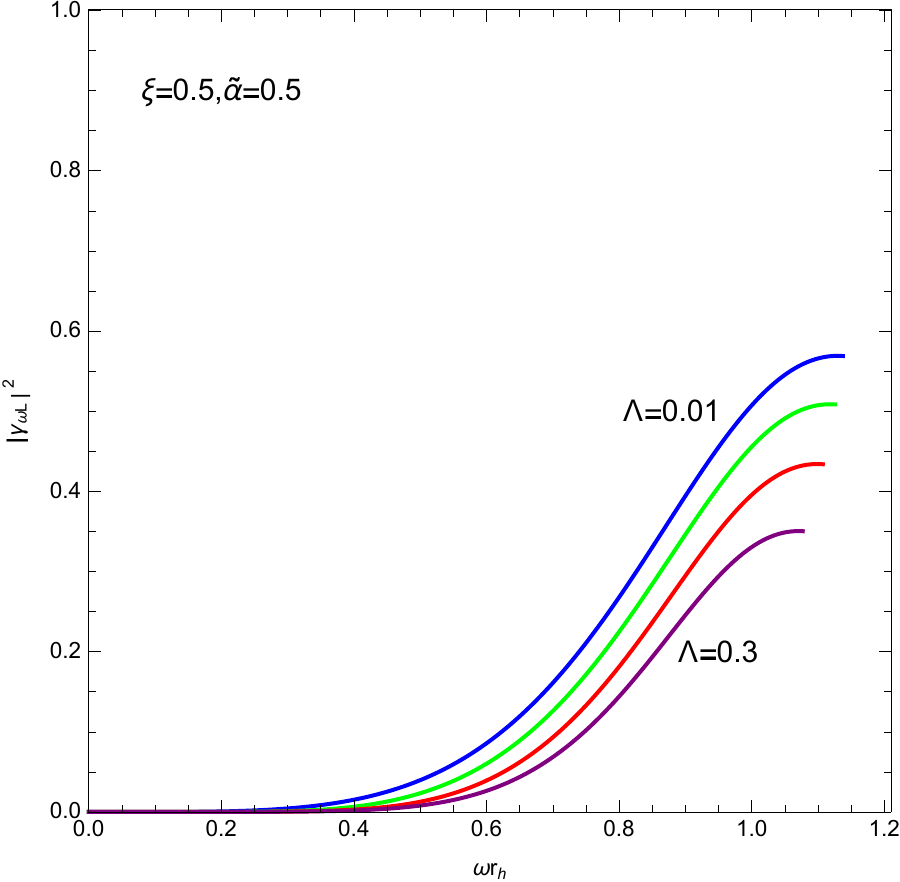}\tabularnewline
\includegraphics[scale=0.6]{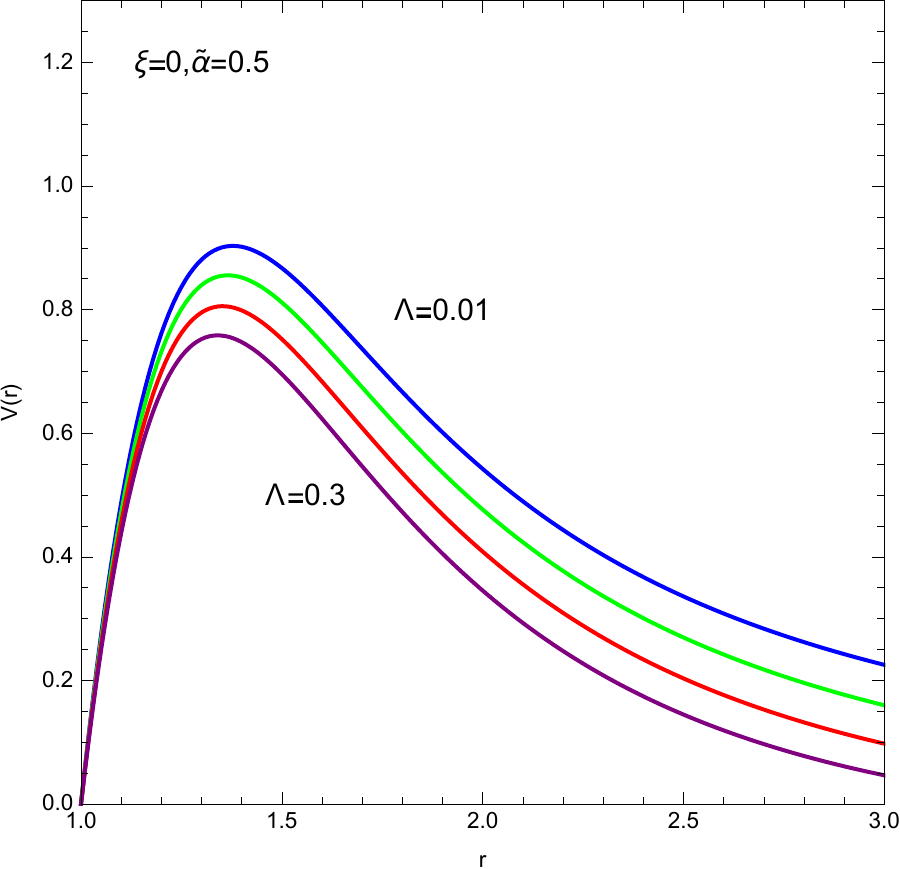} & \includegraphics[scale=0.6]{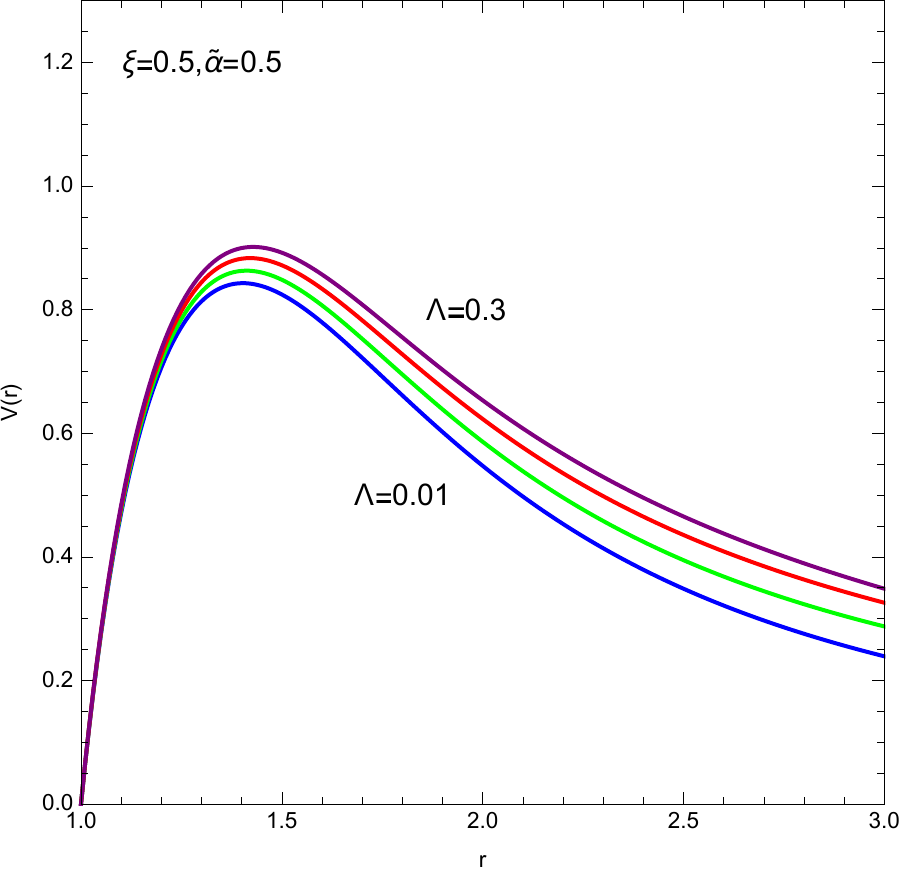}\tabularnewline
\end{tabular}
\par\end{centering}
\caption{\label{fig:GreyLambda} The competition between $\xi$ and $\Lambda$. The greybody factors (upper panel) and corresponding
effective potentials (lower panel) for $d=6,l=0,\tilde{\alpha}=0.5$
with respect to $\Lambda=0.01,0.1,0.2,0.3$. Left panel for $\xi=0$.
Right panel for $\xi=0.5$.}
\end{figure}

 We plot the competition between $\xi$ and $\Lambda$ when $\tilde{\alpha}=0.5$ in Fig. \ref{fig:GreyLambda}.
 As we can see from the left upper panel, when $\xi$ is small, the greybody factor increases with $\Lambda$. However, when $\xi$ is large enough, the greybody factor decreases
with $\Lambda$, as shown in the right upper panel. We show the corresponding
effective potentials in the lower panels. It is obvious that when $\xi$
is small, the potential barrier decreases with $\Lambda$. The situation
is reversed when $\xi$ is large. Thus when $\xi$ is small, $\Lambda$
enhances the greybody factor. When $\xi$ is large enough, $\Lambda$
decreases the greybody factor. The phenomenon is observed  similarly in the SdS case \cite{Kanti2014SdS}. It is due to the double roles
$\Lambda$ plays in the equations of motion. As a homogeneously energy
distributed in the whole spacetime, it subsidizes the energy of emitted
particle and hence enhances the radiation. As an effective mass term
through the non-minimally coupling term, it suppresses the emission. The
competition between these two different contributions leads to the
phenomenon we observed. 

\section{Energy emission rate of Hawking radiation}\label{energyemission}

Greybody factor characterizes the transmissivity of
a particular mode. The more direct quantity is the energy emission
rate, i.e. the power spectra of Hawking radiation. It is given by
\cite{Harris2003,Kanti2004D,Kanti2005}
\begin{equation}
\frac{d^{2}E}{dtd\omega}=\frac{1}{2\pi}\sum_{l}\frac{N_{l}|\gamma_{\omega l}|^{2}\omega}{e^{\omega/T_{BH}}-1}\label{eq:PowerSpectra}
\end{equation}
 where $\omega$ is the energy of the emitted particle, $|\gamma_{\omega l}|^{2}$
the greybody factor in Eq.(\ref{eq:GammaGrey}), $N_{l}=\frac{(2l+d-3)(l+d-4)!}{l!(d-3)!}$
the multiplicity of states that have the same angular momentum number.
$T_{BH}$ is the normalized temperature of the black hole determined
by the surface gravity as \cite{Bousso1996,RKanti2016}
\begin{equation}
T_{BH}=\frac{1}{\sqrt{h(r_{0})}}\frac{1}{4\pi}\left[\frac{(d-2)\left[(d-3)r_{h}^{2}+(d-5)\tilde{\alpha}\right]-2\Lambda r_{h}^{4}}{(d-2)r_{h}(r_{h}^{2}+2\tilde{\alpha})}\right].\label{eq:Temprature}
\end{equation}
 Here $r_{0}$ is the position where $h(r)$ is extreme. We mainly
consider the effects of $\xi$ and $\alpha$ on the power spectra
in this section. Since modes higher than $l>6$ have contributions many orders of magnitude lower than those of the $l\le 6$ modes, their contributions to the energy emission rate are ignored safely.

\subsection{The effects of $\xi$ and $\alpha$}

\begin{figure}[h]
\begin{centering}
\textcolor{red}{}%
\begin{tabular}{cc}
\textcolor{red}{\includegraphics[scale=0.6]{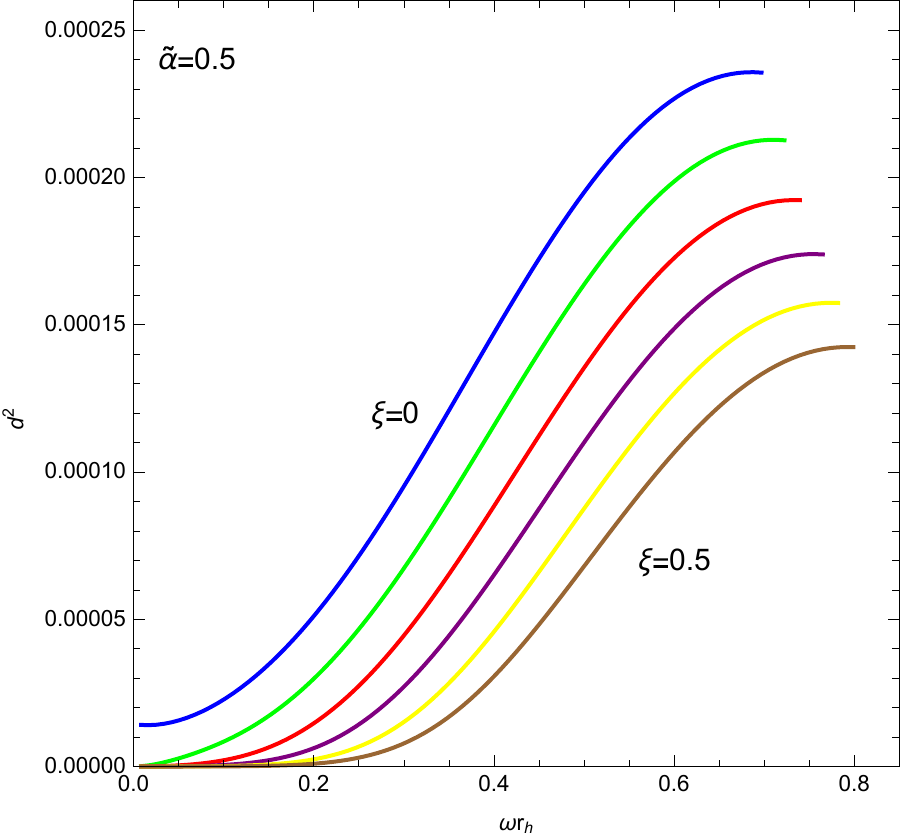}} & \textcolor{red}{\includegraphics[scale=0.6]{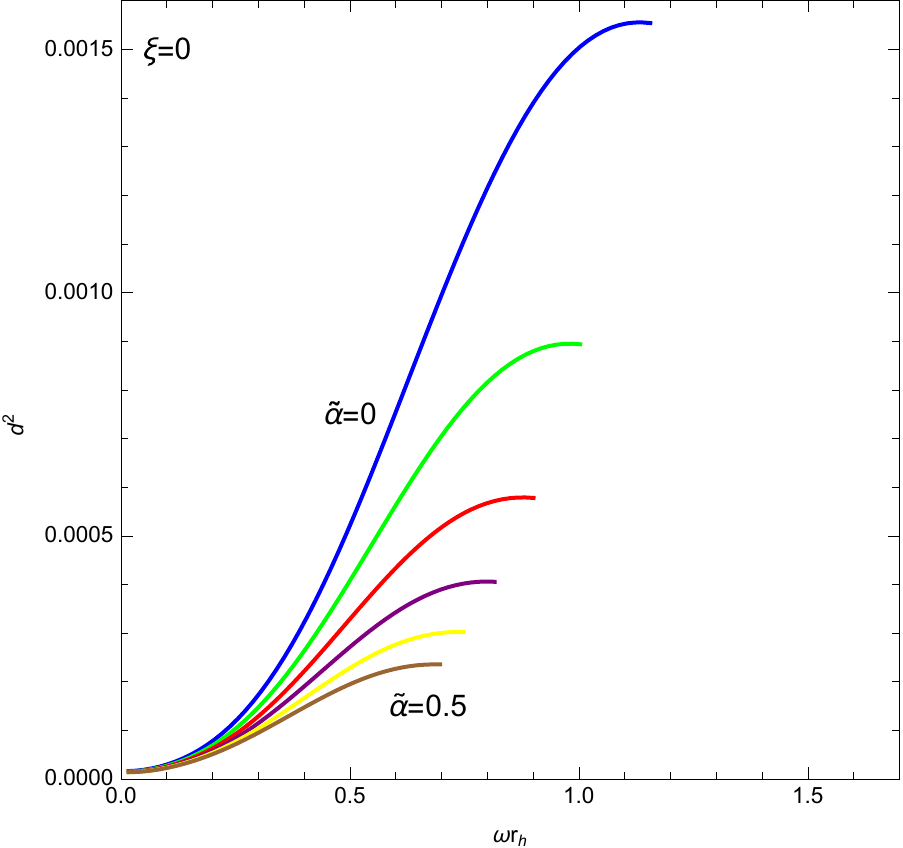}}\tabularnewline
\end{tabular}
\par\end{centering}
\textcolor{red}{\caption{\label{fig:PowerXiAlpha}Power spectra of Hawking radiation for $d=6,\tilde{\alpha}=0.5,\Lambda=0.1$
with respect to $\tilde{\alpha}=0.5$,$\xi=0,0.1,0.2,0.3,0.4,0.5$ (left panel)
and $\xi=0$ and $\tilde{\alpha}=0,0.1,0.2,0.3,0.4,0.5$ (right panel).}
}
\end{figure}

We plot the dependence of power spectra on $\xi$
and $\alpha$ in Fig. \ref{fig:PowerXiAlpha}.

It has been found that $\xi$ suppresses the Hawking
radiation in SdS background. In the left panel, we see that $\xi$
still suppresses the Hawking radiation in EGB-dS background. This
behavior is coincident with that of greybody factor in Fig.\ref{fig:GreyLxi}.
Since the greybody factor decreases with $\xi$, as can be seen from
Eq.(\ref{eq:PowerSpectra}), the power spectra decreases when other
parameters are fixed.

In the right panel, we see that $\tilde{\alpha}$
also suppresses the Hawking radiation. Since $\tilde{\alpha}$ increases
the greybody factor in Fig.\ref{fig:GreyAlpha}, it seems strange
at first sight. However, the power spectra also depends on the temperature
of the black hole. It can be proved easily that the normalized temperature
in Eq.(\ref{eq:Temprature}) decreases with $\tilde{\alpha}$. This
leads to the decrease of the power spectra with $\tilde{\alpha}$
finally.

\textcolor{red}{}

\subsection{The competition between $\xi$ and $\Lambda$}

\begin{figure}[h]
\begin{centering}
\textcolor{red}{}%
\begin{tabular}{cc}
\textcolor{red}{\includegraphics[scale=0.6]{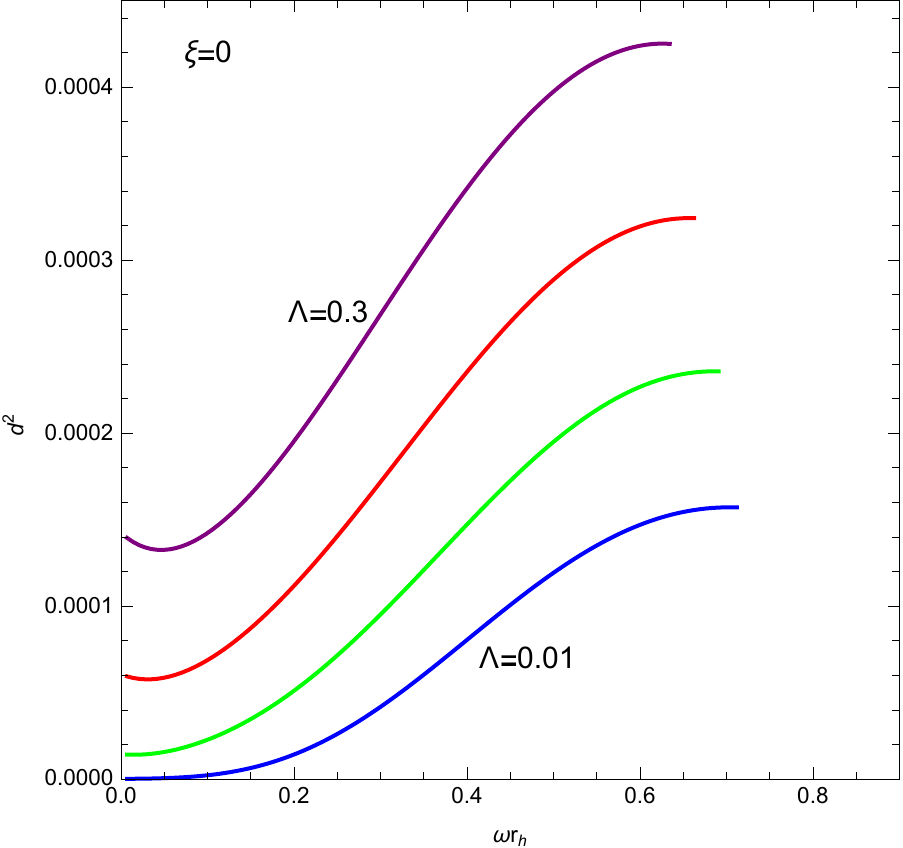}} & \textcolor{red}{\includegraphics[scale=0.6]{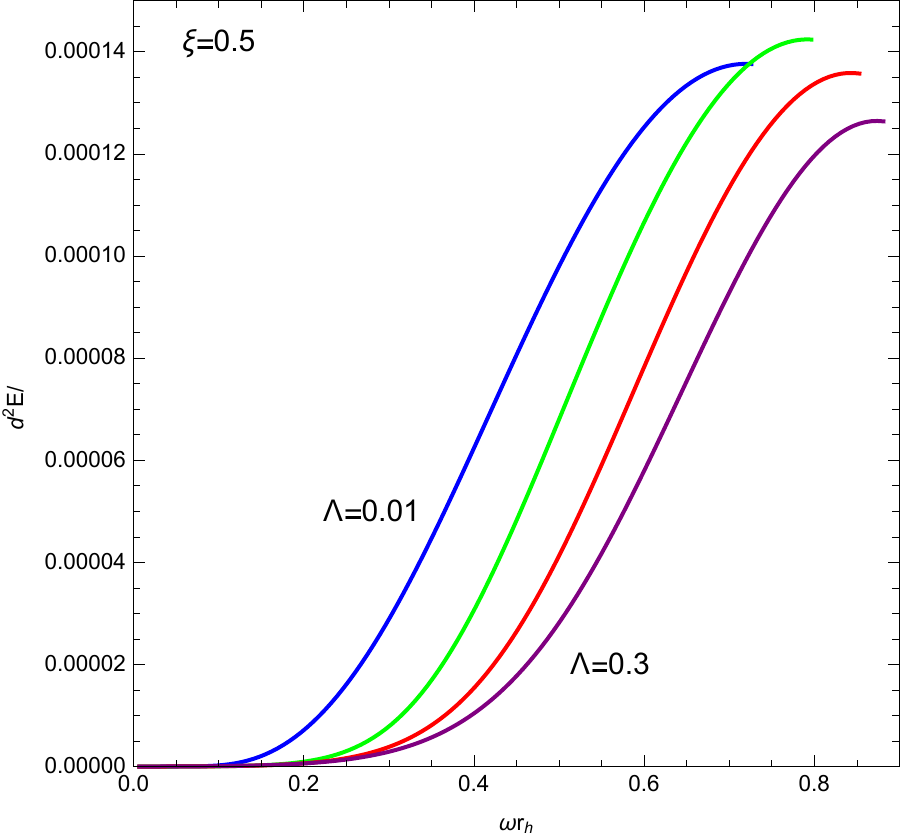}}\tabularnewline
\end{tabular}
\par\end{centering}
\textcolor{red}{\caption{\label{fig:PowerXiLambda}Power spectra of Hawking radiation for $d=6,\tilde{\alpha}=0.5$
with respect to $\Lambda=0.01,0.1,0.2,0.3$. Left panel for $\xi=0$.
Right panel for $\xi=0.5$.}
}
\end{figure}

We have observed that there is a competition between
the contribution of $\xi$ and $\Lambda$ for greybody factor in subsection
\ref{subsec:Competition}. In fact, they have the similar competition
for power spectra of Hawking radiation. We plot their influences on
the power spectra in Fig. \ref{fig:PowerXiLambda}. It is obvious
that when $\xi$ is small, $\Lambda$ increases the Hawking radiation.
When $\xi$ is large enough, $\Lambda$ decreases the Hawking radiation.
Note that the existence of EGB coupling constant does not change this
behavior qualitatively.

\textcolor{red}{}

\section{Conclusion and discussion}\label{conclusion}

We studied the greybody factors of the Hawking radiation for the minimally and
non-minimally coupled scalar fields in a higher dimensional Einstein-Gauss-Bonnet-dS
black hole spacetime. Solving the equations of motion near the event
horizon and cosmological horizon separately and matching them in the
intermediate region, we derived an analytical formula for the greybody
factors when the cosmological constant is small. The larger the distance
between the cosmological horizon and the event horizon, the more accurate
the analytical formula.

The effects of various parameters, such as the angular momentum number
$l$, the non-minimally coupling constant $\xi$, the cosmological constant
$\Lambda$, the GB coupling constant $\tilde{\alpha}$ and the spacetime
dimension $d$, on the greybody factor were studied in detail. We found
that when other parameters are fixed, similar to the case without the GB term, $l,\xi$ or $d$ suppresses
the greybody factor separately. However, the
GB coupling constant $\tilde{\alpha}$ enhances the greybody factor. We
analyzed the competition between $\xi$ and $\tilde{\alpha}$ .
We also studied their effects on the power spectra of Hawking radiation, and found that both of them suppressed the power spectra.
The
effect of the cosmological constant $\Lambda$ is more involved. When
$\xi$ is small, it enhances the greybody factor. When $\xi$ is
large enough, it suppresses the greybody factor. We plotted the
effective potentials to give some intuitive explanations to the phenomenons
we observed.

For the dominant mode $l=0$, the greybody factor for the minimally
coupled scalar is non-vanishing when $\omega=0$. This feature is characteristic for
the free massless scalar propagating in the dS black hole spacetime. For
the EGB-dS black hole, the presence of GB constant $\tilde{\alpha}$ preserves
this feature qualitatively. But quantitatively, it increases the greybody
factor at $\omega=0$.

For the non-minimally coupled scalar, the greybody factors are of order
$O(\omega^{2})$ and vanish for the low energy modes for all the
partial modes $l$ including $l=0$. This can be explained by the
fact that for the non-minimally coupled scalar, $\xi$ plays the role of effective
mass and hinders the Hawking radiation when $\omega\to0$. We obtained
the coefficient of the term at $O(\omega^{2})$ for the EGB-dS black hole
background.

As we mentioned in the context, the results we obtained is only be valid in the low energy region, by using the numerical method we may be able to obtain the  greybody in the  high energy region. We leave this work to future.

\section{Acknowledgments}

We are appreciated Nikalaos Pappas and Panagiota Kanti for their correspondences.
C. Y. Zhang is supported by National Postdoctoral Program for Innovative
Talents BX201600005. B.Chen and P.C. Li were in part supported by NSFC Grant No.~11275010, No.~11325522 and
No. 11735001.

\end{document}